\documentclass[12pt,aps,prl,showpacs,superscriptaddress,preprint,byrevtex]{revtex4}

\usepackage{rotating}
\usepackage{longtable}
\usepackage{amsmath}
\usepackage{graphicx} 
\usepackage{dcolumn}  
\usepackage{multirow}

\renewcommand{\arraystretch}{1.1}

\newcommand{\cm}{\,\mathrm{cm}}
\newcommand{\mm}{\,\mathrm{mm}}
\newcommand{\mev}{\mathrm{MeV}}

\newcommand{\mevm}{\mathrm{MeV}/c^2}
\newcommand{\gev}{\mathrm{GeV}}

\newcommand{\gevm}{\mathrm{GeV}/c^2}
\newcommand{\gevms}{\mathrm{GeV}^2/c^4}

\newcommand{\ee}{e^+e^-}
\newcommand{\uu}{\mu^+\mu^-}
\newcommand{\pp}{\pi^+\pi^-}
\newcommand{\U}{\Upsilon}
\newcommand{\Uf}{\Upsilon(5S)}
\newcommand{\Uo}{\Upsilon(1S)}
\newcommand{\Un}{\Upsilon(nS)}
\newcommand{\Ut}{\Upsilon(2S)}
\newcommand{\Uth}{\Upsilon(3S)}

\newcommand{\mmpp}{MM(\pi^+\pi^-)}
\newcommand{\mmp}{MM(\pi)}
\newcommand{\mmpip}{MM(\pi^+)}
\newcommand{\mmpim}{MM(\pi^-)}
\newcommand{\mpp}{M(\pi^+\pi^-)}
\newcommand{\hb}{h_b(1P)}

\newcommand{\hbp}{h_b(2P)}
\newcommand{\hbn}{h_b(mP)}
\newcommand{\ks}{K^0_S}
\newcommand{\pip}{\pi^{+}}
\newcommand{\pipm}{\pi^{\pm}}
\newcommand{\pim}{\pi^{-}}
\newcommand{\fb}{\mathrm{fb}^{-1}}

\newcommand{\etal}{\em et al.}

\newcommand{\zb}{Z_b}
\newcommand{\zbo}{Z_b(10610)}
\newcommand{\zbt}{Z_b(10650)}
\newcommand{\cto}{\cos\theta_1}
\newcommand{\ctt}{\cos\theta_2}
\newcommand{\ctpp}{\cos\theta_{\pi\pi}}

\newcommand{\mzahb}{10605.1\pm2.2\,^{+3.0}_{-1.0}}
\newcommand{\gzahb}{11.4\,^{+4.5}_{-3.9}\,^{+2.1}_{-1.2}}
\newcommand{\mzbhb}{10654.5\pm2.5\,^{+1.0}_{-1.9}}
\newcommand{\gzbhb}{20.9\,^{+5.4}_{-4.7}\,^{+2.1}_{-5.7}}
\newcommand{\ahb}{1.8\,^{+1.0}_{-0.7}\,^{+0.1}_{-0.5}}
\newcommand{\phihb}{188\,^{+44}_{-58}\,^{+4}_{-9}}

\newcommand{\mzahbp}{10596\pm7\,^{+5}_{-2}}
\newcommand{\gzahbp}{16\,^{+16}_{-10}\,^{+13}_{-4}}
\newcommand{\mzbhbp}{10651\pm4\pm2}
\newcommand{\gzbhbp}{12\,^{+11}_{-9}\,^{+8}_{-2}}
\newcommand{\ahbp}{1.3\,^{+3.1}_{-1.1}\,^{+0.4}_{-0.7}}
\newcommand{\phihbp}{255\,^{+56}_{-72}\,^{+12}_{-183}}

\begin{document}

\title{\boldmath Observation of two charged bottomonium-like resonances}

\date{\today}

\begin{abstract}
\noindent
We report the observation of two narrow structures at $10610\,\mevm$ and
$10650\,\mevm$ in the $\pi^\pm\Un$ ($n=1,2,3$) and $\pi^\pm\hbn$
($m=1,2$) mass spectra that are produced in association with a single
charged pion in $\Uf$ decays. The measured masses and widths of the
two structures averaged over the five final states are
$M_1=10608.4\pm2.0\,\mevm$,
$\Gamma_1=15.6\pm2.5\,\mev$ and
$M_2=10653.2\pm1.5\,\mevm$,
$\Gamma_2=14.4\pm3.2\,\mev$.  Analysis favors quantum
numbers of $I^{G}(J^{P})$=$1^{+}(1^{+})$ for both states. The results
are obtained with a $121.4\,{\rm fb}^{-1}$ data sample collected with
the Belle detector near the $\Uf$ resonance at the KEKB
asymmetric-energy $\ee$ collider.
\end{abstract}


\affiliation{University of Bonn, Bonn}
\affiliation{Budker Institute of Nuclear Physics, Novosibirsk}
\affiliation{Faculty of Mathematics and Physics, Charles University, Prague}
\affiliation{Chiba University, Chiba}
\affiliation{University of Cincinnati, Cincinnati, Ohio 45221}
\affiliation{Department of Physics, Fu Jen Catholic University, Taipei}
\affiliation{Justus-Liebig-Universit\"at Gie\ss{}en, Gie\ss{}en}
\affiliation{Gifu University, Gifu}
\affiliation{The Graduate University for Advanced Studies, Hayama}
\affiliation{Gyeongsang National University, Chinju}
\affiliation{Hanyang University, Seoul}
\affiliation{University of Hawaii, Honolulu, Hawaii 96822}
\affiliation{High Energy Accelerator Research Organization (KEK), Tsukuba}
\affiliation{Hiroshima Institute of Technology, Hiroshima}
\affiliation{University of Illinois at Urbana-Champaign, Urbana, Illinois 61801}
\affiliation{Indian Institute of Technology Guwahati, Guwahati}
\affiliation{Indian Institute of Technology Madras, Madras}
\affiliation{Indiana University, Bloomington, Indiana 47408}
\affiliation{Institute of High Energy Physics, Chinese Academy of Sciences, Beijing}
\affiliation{Institute of High Energy Physics, Vienna}
\affiliation{Institute of High Energy Physics, Protvino}
\affiliation{Institute of Mathematical Sciences, Chennai}
\affiliation{INFN - Sezione di Torino, Torino}
\affiliation{Institute for Theoretical and Experimental Physics, Moscow}
\affiliation{J. Stefan Institute, Ljubljana}
\affiliation{Kanagawa University, Yokohama}
\affiliation{Institut f\"ur Experimentelle Kernphysik, Karlsruher Institut f\"ur Technologie, Karlsruhe}
\affiliation{Korea Institute of Science and Technology Information, Daejeon}
\affiliation{Korea University, Seoul}
\affiliation{Kyoto University, Kyoto}
\affiliation{Kyungpook National University, Taegu}
\affiliation{\'Ecole Polytechnique F\'ed\'erale de Lausanne (EPFL), Lausanne}
\affiliation{Faculty of Mathematics and Physics, University of Ljubljana, Ljubljana}
\affiliation{Luther College, Decorah, Iowa 52101}
\affiliation{University of Maribor, Maribor}
\affiliation{Max-Planck-Institut f\"ur Physik, M\"unchen}
\affiliation{University of Melbourne, School of Physics, Victoria 3010}
\affiliation{Nagoya University, Nagoya}
\affiliation{Nara University of Education, Nara}
\affiliation{Nara Women's University, Nara}
\affiliation{National Central University, Chung-li}
\affiliation{National United University, Miao Li}
\affiliation{Department of Physics, National Taiwan University, Taipei}
\affiliation{H. Niewodniczanski Institute of Nuclear Physics, Krakow}
\affiliation{Nippon Dental University, Niigata}
\affiliation{Niigata University, Niigata}
\affiliation{University of Nova Gorica, Nova Gorica}
\affiliation{Novosibirsk State University, Novosibirsk}
\affiliation{Osaka City University, Osaka}
\affiliation{Osaka University, Osaka}
\affiliation{Pacific Northwest National Laboratory, Richland, Washington 99352}
\affiliation{Panjab University, Chandigarh}
\affiliation{Peking University, Beijing}
\affiliation{Princeton University, Princeton, New Jersey 08544}
\affiliation{Research Center for Nuclear Physics, Osaka}
\affiliation{RIKEN BNL Research Center, Upton, New York 11973}
\affiliation{Saga University, Saga}
\affiliation{University of Science and Technology of China, Hefei}
\affiliation{Seoul National University, Seoul}
\affiliation{Shinshu University, Nagano}
\affiliation{Sungkyunkwan University, Suwon}
\affiliation{School of Physics, University of Sydney, NSW 2006}
\affiliation{Tata Institute of Fundamental Research, Mumbai}
\affiliation{Excellence Cluster Universe, Technische Universit\"at M\"unchen, Garching}
\affiliation{Toho University, Funabashi}
\affiliation{Tohoku Gakuin University, Tagajo}
\affiliation{Tohoku University, Sendai}
\affiliation{Department of Physics, University of Tokyo, Tokyo}
\affiliation{Tokyo Institute of Technology, Tokyo}
\affiliation{Tokyo Metropolitan University, Tokyo}
\affiliation{Tokyo University of Agriculture and Technology, Tokyo}
\affiliation{Toyama National College of Maritime Technology, Toyama}
\affiliation{CNP, Virginia Polytechnic Institute and State University, Blacksburg, Virginia 24061}
\affiliation{Wayne State University, Detroit, Michigan 48202}
\affiliation{Yamagata University, Yamagata}
\affiliation{Yonsei University, Seoul}
  \author{I.~Adachi}\affiliation{High Energy Accelerator Research Organization (KEK), Tsukuba} 
  \author{K.~Adamczyk}\affiliation{H. Niewodniczanski Institute of Nuclear Physics, Krakow} 
  \author{H.~Aihara}\affiliation{Department of Physics, University of Tokyo, Tokyo} 
  \author{K.~Arinstein}\affiliation{Budker Institute of Nuclear Physics, Novosibirsk}\affiliation{Novosibirsk State University, Novosibirsk} 
  \author{Y.~Arita}\affiliation{Nagoya University, Nagoya} 
  \author{D.~M.~Asner}\affiliation{Pacific Northwest National Laboratory, Richland, Washington 99352} 
  \author{T.~Aso}\affiliation{Toyama National College of Maritime Technology, Toyama} 
  \author{V.~Aulchenko}\affiliation{Budker Institute of Nuclear Physics, Novosibirsk}\affiliation{Novosibirsk State University, Novosibirsk} 
  \author{T.~Aushev}\affiliation{\'Ecole Polytechnique F\'ed\'erale de Lausanne (EPFL), Lausanne}\affiliation{Institute for Theoretical and Experimental Physics, Moscow} 
  \author{T.~Aziz}\affiliation{Tata Institute of Fundamental Research, Mumbai} 
% ...
\collaboration{The Belle Collaboration}

\maketitle

{\renewcommand{\thefootnote}{\fnsymbol{footnote}}}
\setcounter{footnote}{0}

\section{Introduction}

Analyses of decays of the $\Uf$ resonance to non-$B_s\bar{B}_s$ final states
have produced several surprises. Recently the Belle Collaboration reported
observation of anomalously high rates for $\Uf\to\Un\pp$ 
($n=1,2,3$)~\cite{Belle_ypipi} and $\Uf\to\hbn\pp$ ($m=1,2$)~\cite{Belle_hb}
transitions. If the $\Un$ signals are attributed entirely to $\Uf$ decays,
the measured partial decay widths $\Gamma[\Uf\to\Un\pp]\sim0.5\,\mev$ are 
about two orders of magnitude larger than typical widths for dipion transitions
among the four lower $\U(nS)$ states. The rates for the $\Uf\to\hbn\pp$
transitions are found to be comparable with those for $\Uf\to\Un\pp$, 
indicating that processes requiring a heavy-quark spin flip - {i.e.,} in
$\hbn$ production - do not seem to be suppressed. These unexpected observations
indicate that exotic mechanisms could be contributing to the $\Uf$ decays.

We report preliminary results of detailed studies of the three-body 
$\Uf\to\Un\pp$ ($n=1,2,3$) and $\Uf\to\hbn\pp$ ($m=1,2$) decays. 
Results are obtained with a $121.4\,\fb$ data sample collected near
the peak of the $\Uf$ resonance ($\sqrt{s}\sim 10.865\gev$) with the
Belle detector at the KEKB asymmetric energy $\ee$
collider~\cite{KEKB}.

\section{Belle Detector}

The Belle detector is a large-solid-angle magnetic spectrometer that
consists of a 4-layer silicon vertex detector (SVD), a 50-layer
central drift chamber (CDC), an array of aerogel threshold Cherenkov
counters (ACC), a barrel-like arrangement of time-of-flight
scintillation counters (TOF), and an electromagnetic calorimeter (ECL)
comprising CsI(Tl) crystals located inside a superconducting solenoid
coil that provides a 1.5$\,$T magnetic field.  An iron flux-return
located outside of the coil is instrumented to detect $K^0_L$ mesons
and to identify muons (KLM). The detector is described in detail
elsewhere~\cite{BELLE_DETECTOR}.

For charged hadron identification, the $dE/dx$ measurement in the CDC
and the response of the ACC and TOF are combined to form a single
likelihood ratio. Electron identification is based on a combination of
$dE/dx$ measurement, the response of the ACC, and the position, shape
and total energy deposition of the shower detected in the ECL. Muons
are identified by their range and transverse scattering in the KLM.

We use a GEANT-based Monte Carlo (MC) simulation~\cite{geant} to model
the response of the detector and determine the acceptance. The MC
simulation includes run-dependent detector performance variations and
background conditions.

\section{\boldmath Analysis of $\Upsilon(5S)\to\Upsilon(1S,2S,3S)\pi^+\pi^-$}

To select $\Uf\to\Un\pp$ candidate events we require the presence of a pair
of muon candidates with an invariant mass in the range of 
$8.0~\gevm<M(\mu^+\mu^-)<11.0~\gevm$ and two pion candidates of opposite
charge. These tracks are required to be consistent with coming from the
interaction point: we apply requirements $dr<0.5$~cm and $|dz|<2.5$~cm,
where $dr$ and $|dz|$ are impact parameters perpendicular to and along the
beam axis with respect to the interaction point, respectively. We also
require that none of the four tracks is positively identified as an electron.
No additional requirements are applied at this stage.

\begin{figure}[!t]
  \centering
\hspace*{-1mm}
  \includegraphics[width=0.48\textwidth]{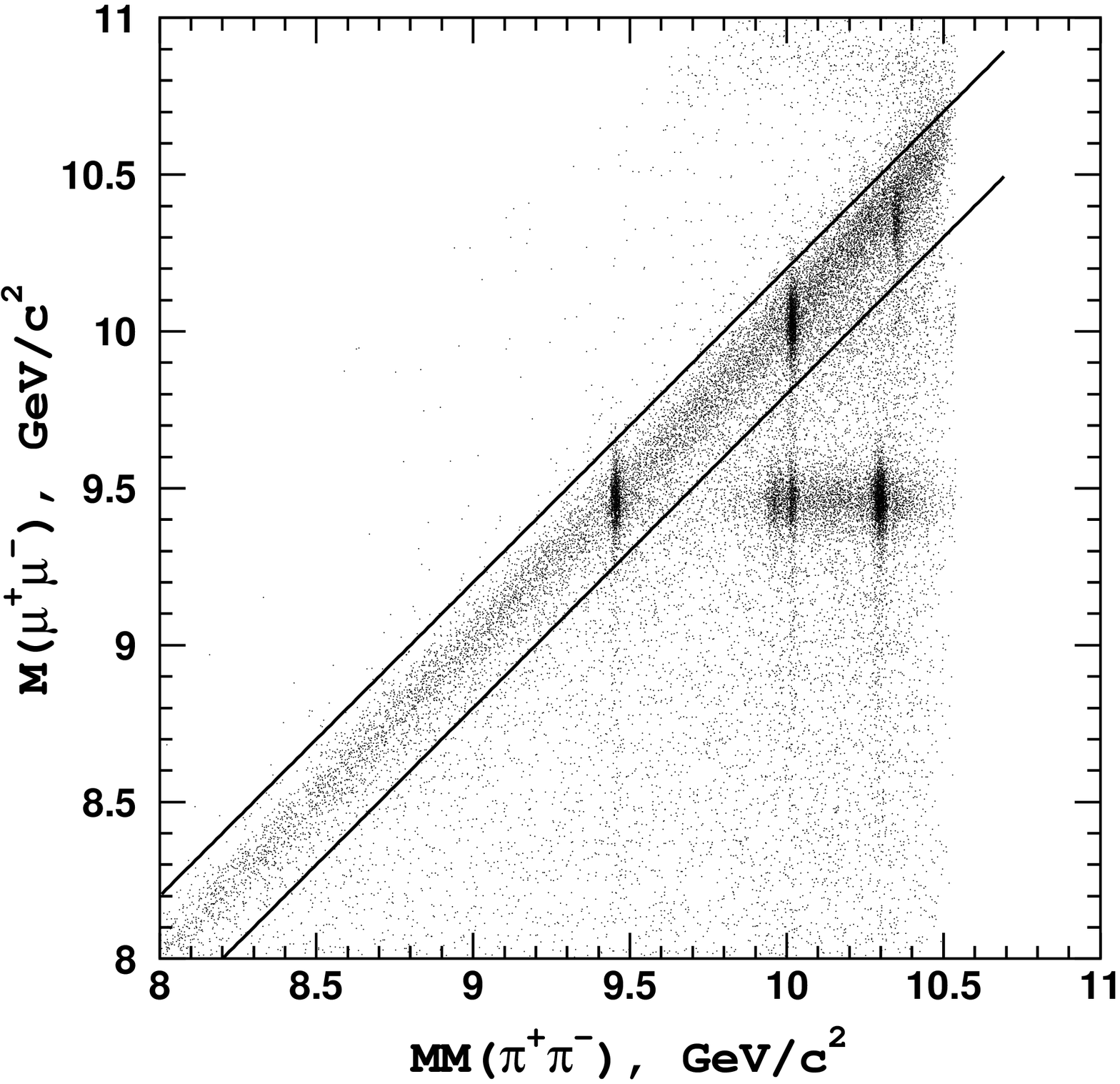}
  \caption{A scatter plot of all of the $\Uf\to\Un\pp$
           candidate events that pass the initial selection criteria.
           The area between the two diagonal lines is the $\Uf$ signal
           region.}
\label{fig:ynspp-s-y5}
\end{figure}

Candidate $\Uf\to\Un\pp$ events are identified by the invariant mass of the
$\mu^+\mu^-$ combination and the missing mass $MM(\pi^+\pi^-)$ associated
with the $\pp$ system calculated as
\begin{equation}
MM(\pp) = \sqrt{(E_{c.m.}-E_{\pp}^*)^2-p_{\pp}^{*2}},
\end{equation}
where $E_{\rm c.m.}$ is the center-of-mass (c.m.) energy and $E^*_{\pp}$ 
and $p^*_{\pp}$ are the energy and momentum of the $\pp$ system measured
in the c.m.\ frame. The two-dimensional distribution of $M(\mu^+\mu^-)$
versus $MM(\pp)$ for all the preselected candidates is shown in 
Fig.~\ref{fig:ynspp-s-y5}. Events originated from $\Upsilon(5S)$ decay 
fall within the narrow diagonal band ($\Uf$ signal region) defined as 
$|MM(\pp)-M(\mu^+\mu^-)|<0.2$~GeV/$c^2$ (see Fig.~\ref{fig:ynspp-s-y5}).
Clustering of events around nominal $\Un$ mass values~\cite{PDG} are clearly
visible on the plot.

The amplitude analyses of the three-body $\Uf\to\Un\pp$ decays that are
reported here are performed by means of unbinned maximum likelihood fits to
two-dimensional Dalitz distributions.

\begin{figure}[!t]
  \centering
\hspace*{-1mm}
  \includegraphics[width=0.32\textwidth]{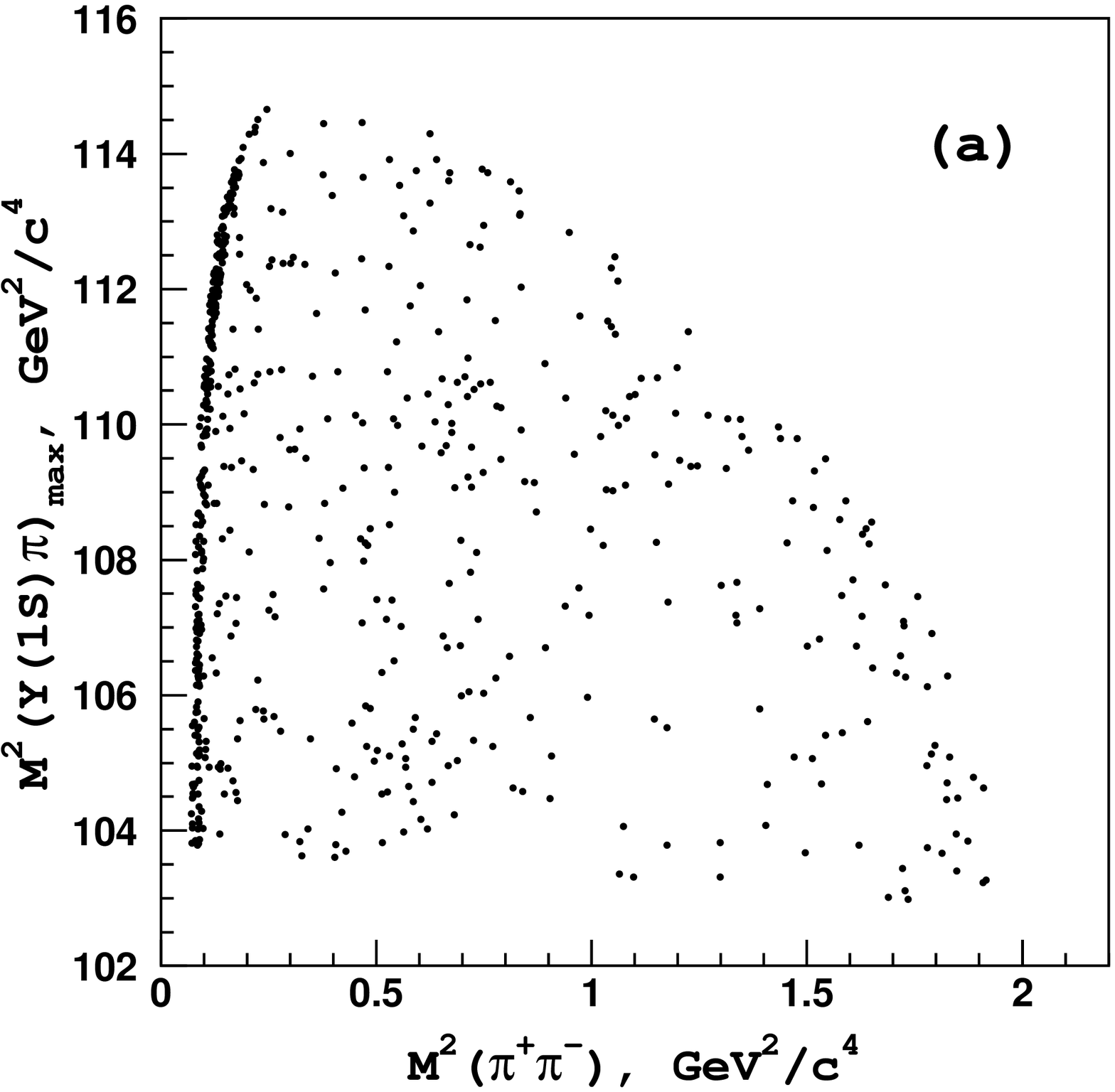} \hfill
  \includegraphics[width=0.32\textwidth]{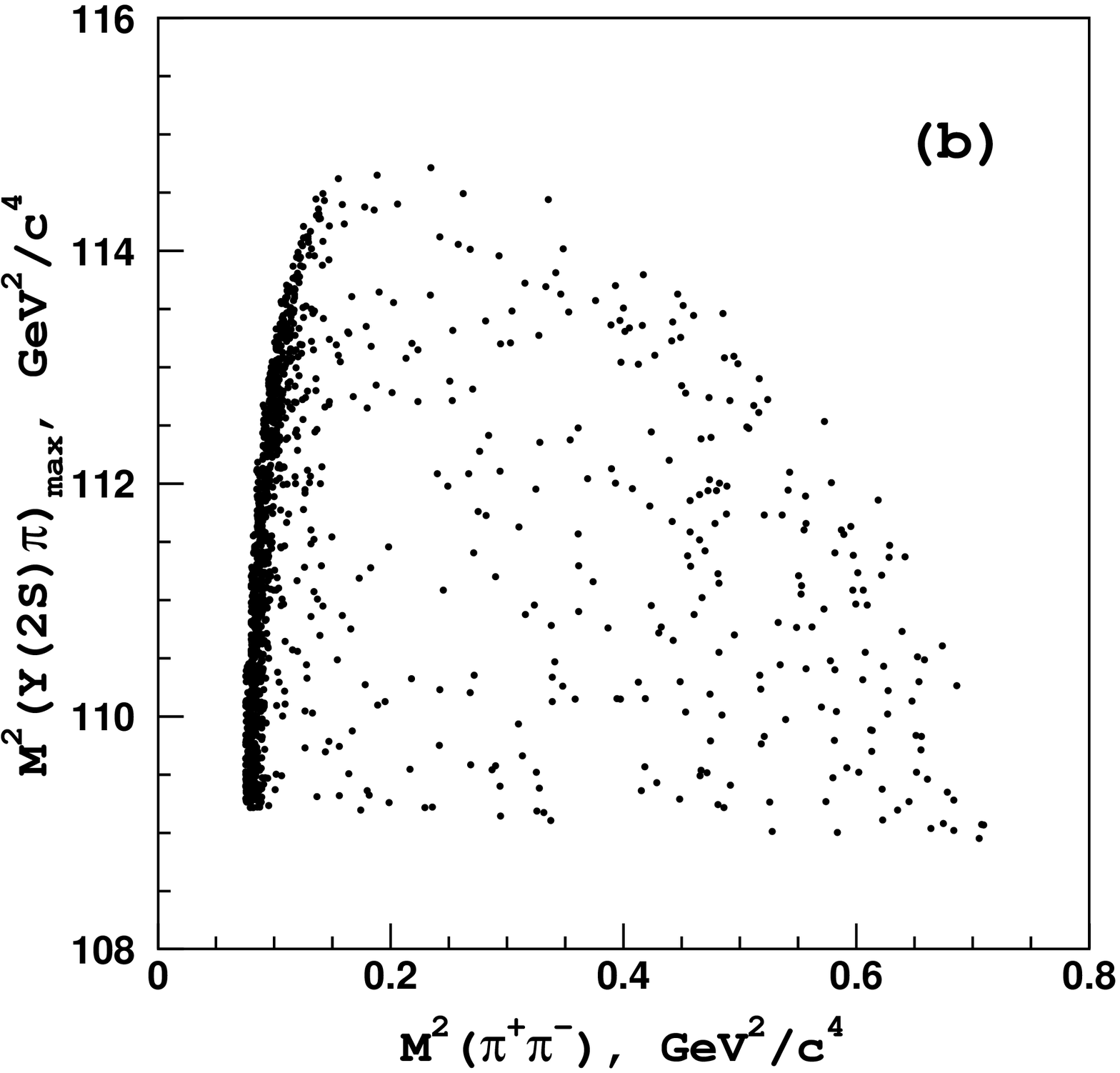} \hfill
  \includegraphics[width=0.32\textwidth]{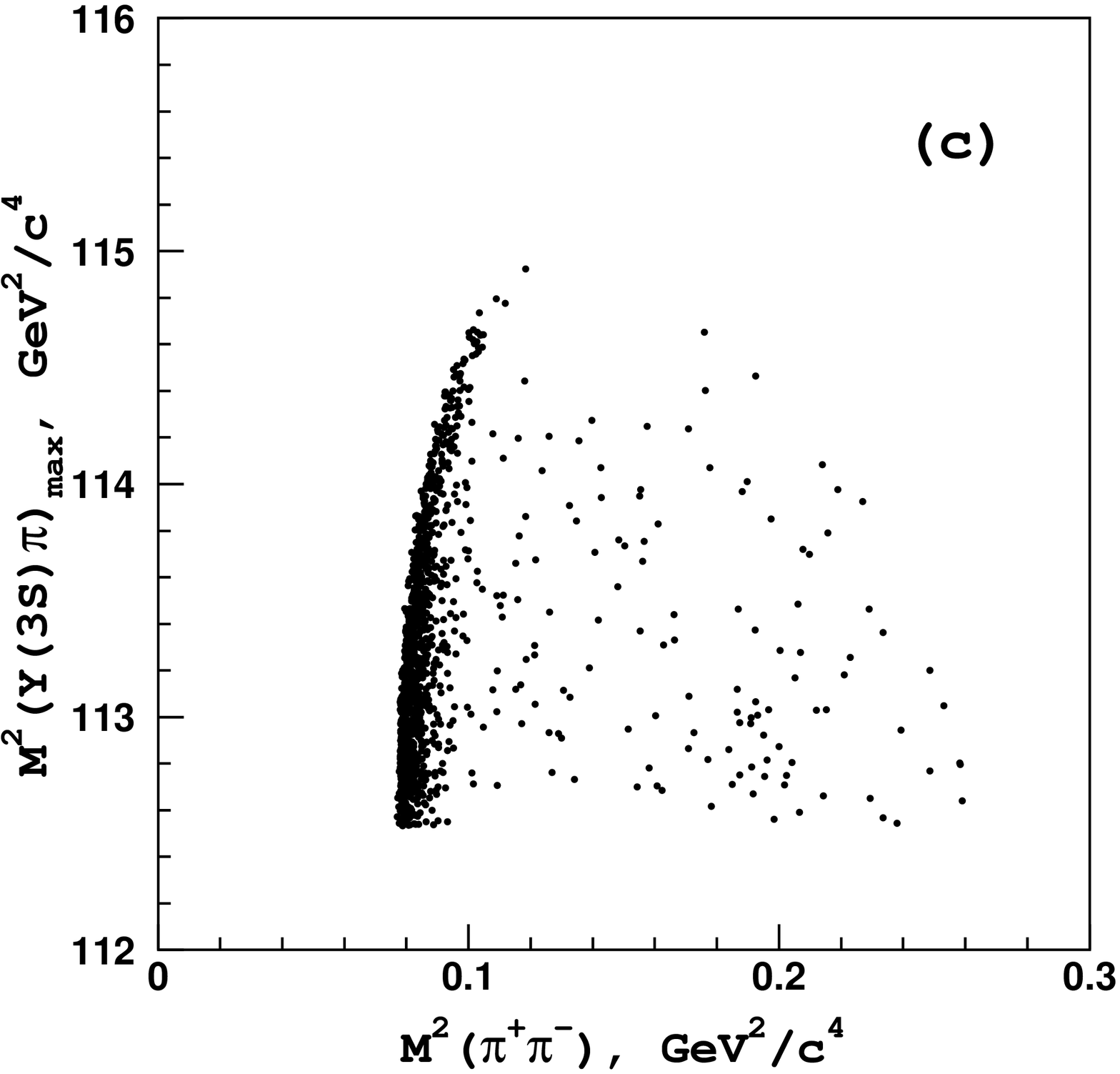}
  \caption{Dalitz plots for $\Un\pp$ events in the (a) $\Uo$; (b)
    $\Ut$; \mbox{(c) $\Uth$} sidebands. }
\label{fig:ynspp-b-dp}
\end{figure}

Before fitting the Dalitz plot for events in the signal region, we determine
the distribution of background events over the Dalitz plot using events in
the $\Upsilon(nS)$ mass sidebands that are refitted to the nominal mass
of the corresponding $\Un$ state to match the phase space boundaries. 
Definitions of the mass sidebands are given in Table~\ref{tab:ynspp_sides} 
and the corresponding sideband Dalitz plots are shown in 
Fig.~\ref{fig:ynspp-b-dp}, where $M(\Un\pi)_{\min}$ and $M(\Un\pi)_{\max}$
are defined as $M(\Un\pi)_{\min}=\min(M(\Un\pi^+),M(\Un\pi^-))$ and
$M(\Un\pi)_{\max}=\max(M(\Un\pi^+),M(\Un\pi^-))$, respectively.

\begin{table}[!b]
  \caption{Definition of the sidebands for $\Un\pp$ signals.  The
    sidebands are defined in terms of regions in the spectrum of the
    missing mass associated with the $\pp$ system.}
  \medskip
  \label{tab:ynspp_sides}
\centering
  \begin{tabular}{lccc} \hline \hline
Final state & ~$\Uo\pp$~       &
              ~$\Ut\pp$~       &
              ~$\Uth\pp$~
\\ \hline
   Sideband &
               ~$ 9.38<MM(\pi\pi)< 9.43$~    &
               ~$ 9.95<MM(\pi\pi)<10.00$~    &
               ~$10.28<MM(\pi\pi)<10.33$~    
\vspace*{-3mm} \\
     region, GeV/$c^2$ &
               ~$ 9.48<MM(\pi\pi)< 9.53$~    &
               ~$10.05<MM(\pi\pi)<10.10$~    &
               ~$10.38<MM(\pi\pi)<10.43$~    
\vspace*{1mm} \\
\hline \hline
\end{tabular}
\end{table}

\begin{figure}[!t]
  \centering
\hspace*{-1mm}
  \includegraphics[width=0.32\textwidth]{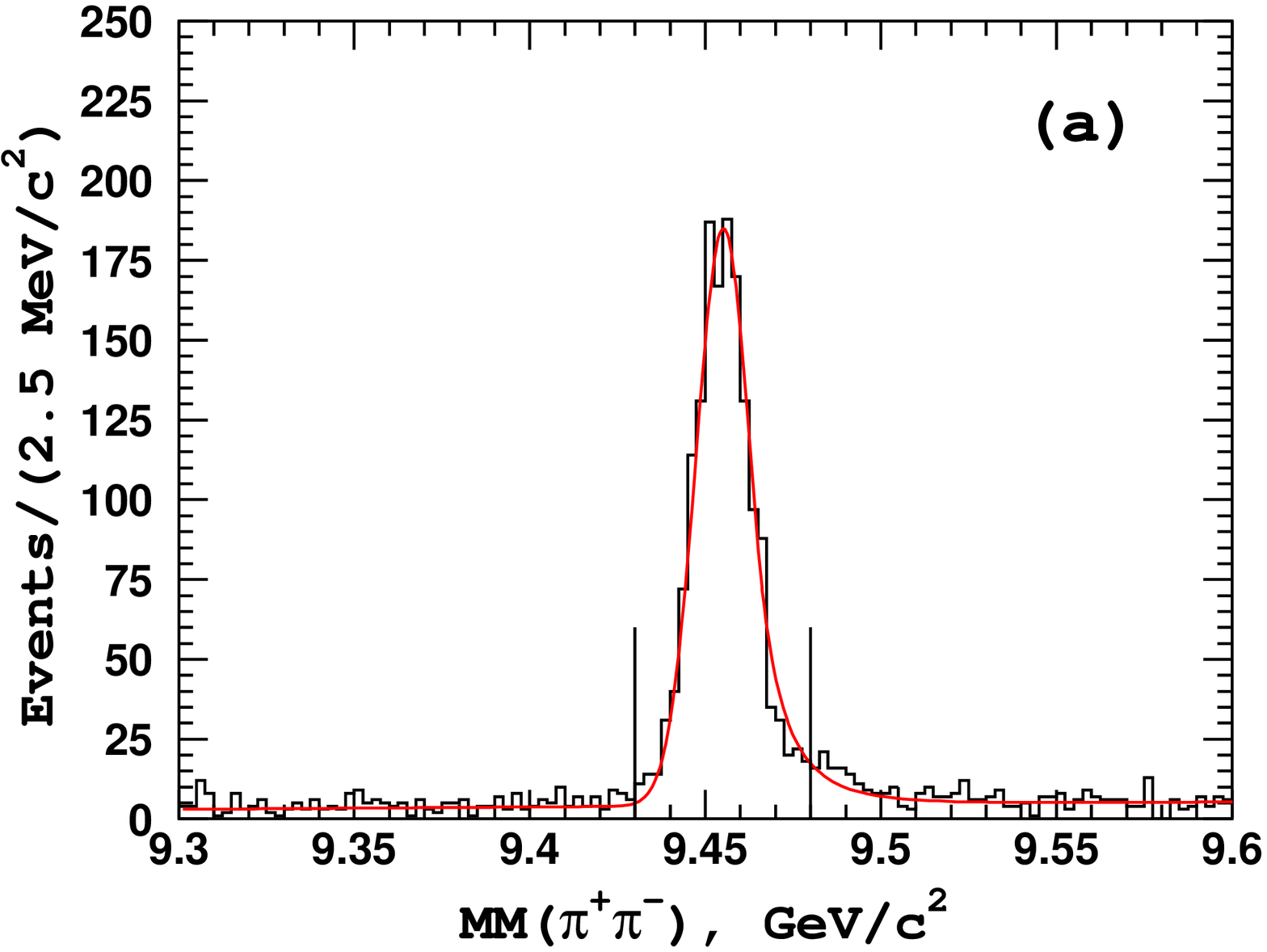} \hfill
  \includegraphics[width=0.32\textwidth]{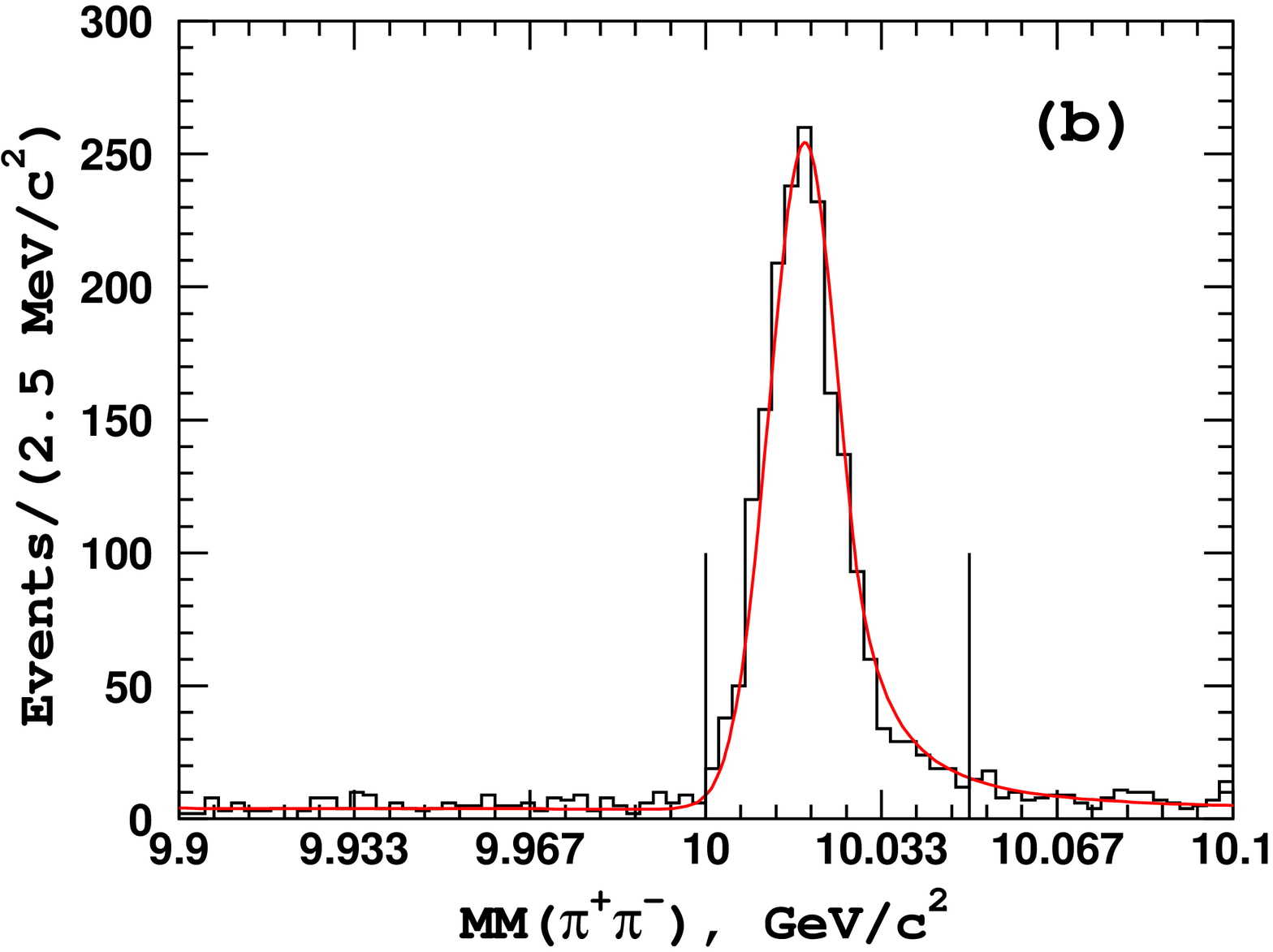} \hfill
  \includegraphics[width=0.32\textwidth]{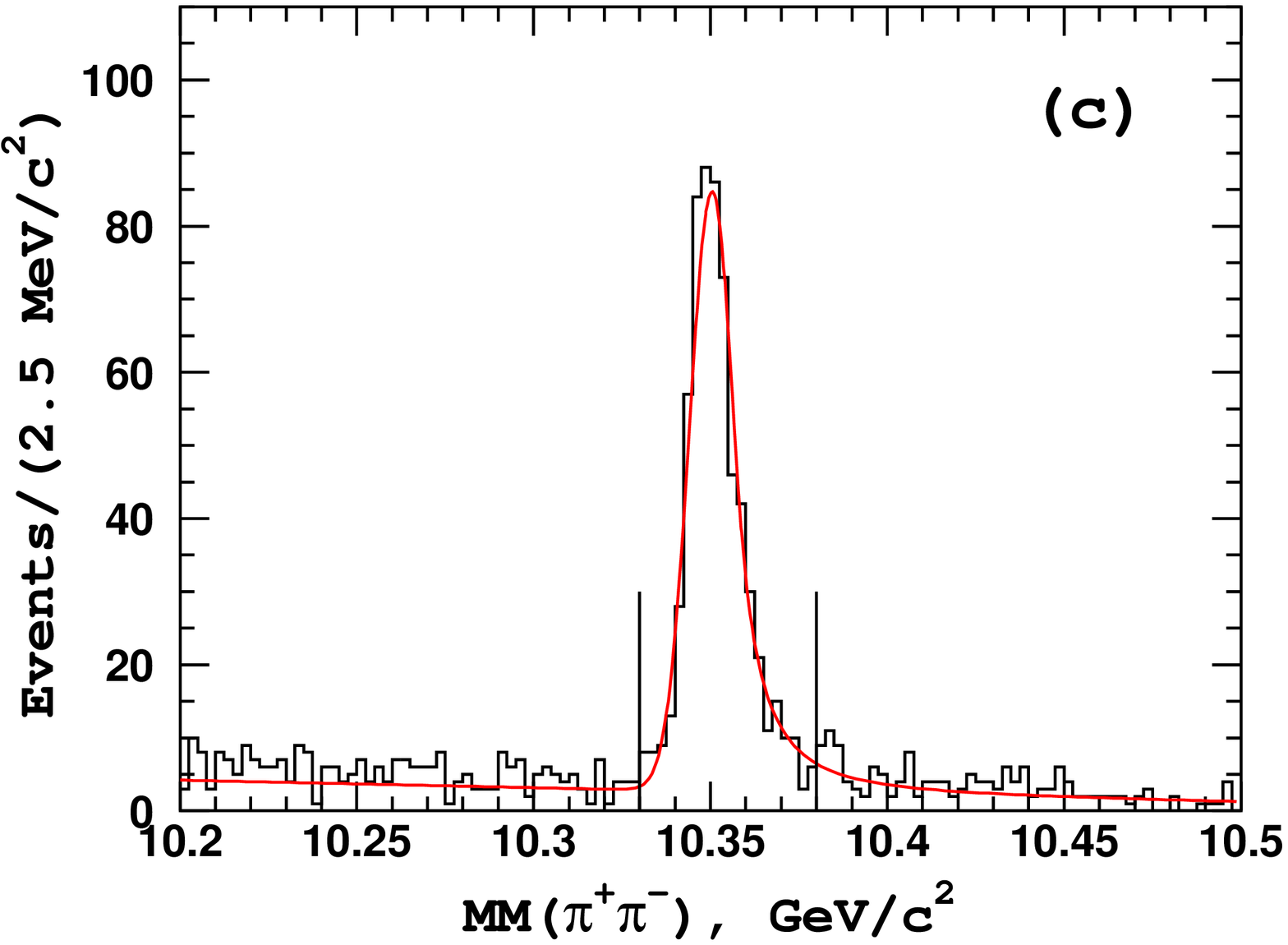}
  \caption{Distribution of missing mass associated with the $\pp$ 
           combination for $\Uf\to\Un\pp$ candidate events in the
           (a) $\Uo$; (b) $\Ut$; (c) $\Uth$ mass regions. 
           Vertical lines define the corresponding signal regions.}
\label{fig:ynspp-s-mm}
\end{figure}

It is evident in the sideband Dalitz distributions, shown in 
Fig.~\ref{fig:ynspp-b-dp}, that there is a strong concentration of 
background events in the very low $\pp$ invariant mass region; these are
due to photon conversion in the innermost parts of the Belle detector. 
Because of their low energy, these conversion electrons are poorly 
identified and pass the electron veto requirement. We exclude this high
background region by applying the requirements on the $\pp$ invariant
mass given in Table~\ref{tab:ynspp_sfrac}. For the remainder of the
Dalitz plot the distribution of background events is assumed to
be uniform. The variation of reconstruction efficiency across the Dalitz
plot is determined from MC simulation. The fraction of signal events in the
signal region for each of the three $\Un\pp$ final states is determined 
from a fit to the corresponding $MM(\pp)$ spectrum using a Crystal Ball
function~\cite{CBF} for the $\Upsilon$ signal and a linear function for
the combinatorial background component. Results of the fits are shown in
Fig.~\ref{fig:ynspp-s-mm} and are summarized in Table~\ref{tab:ynspp_sfrac}.

\begin{table}[!b]
  \caption{Results of the fits to the $MM(\pi^+\pi^-)$ missing mass
    distributions.}
  \medskip
  \label{tab:ynspp_sfrac}
\centering
  \begin{tabular}{lccc} \hline \hline
Final state & ~$\Uo\pp$~   &
              ~$\Ut\pp$~   &
              ~$\Uth\pp$~
\\ \hline
            $M(\pp)$ signal   &
         ~~$0.20<M(\pp)$~~    &
         ~~$0.16<M(\pp)$~~    &
         ~~$0.10<M(\pp)$~~    
\vspace*{-3mm} \\
            region, GeV/$c^2$   &
               &
               &
\vspace*{0mm} \\
            $MM(\pp)$ signal    &
         ~~$ 9.43<MM(\pp)$~~    &
         ~~$10.00<MM(\pp)$~~    &
         ~~$10.33<MM(\pp)$~~    
\vspace*{-3mm} \\
            region, GeV/$c^2$   &
         ~$MM(\pp)< 9.48$~~    &
         ~$MM(\pp)<10.05$~~    &
         ~$MM(\pp)<10.38$~~    
\vspace*{0mm} \\
            Peak position,        &
           $9455.3\pm1.1$                    &
           $10019.1\pm1.7$                    &
           $10350.2\pm2.3$    
\vspace*{-3mm} \\
             GeV/$c^2$    &
               &
               &
\vspace*{0mm} \\
            Resolution,            &
           $7.31$                            &
           $6.74$                            &
           $6.33$    
\vspace*{-3mm} \\
             MeV/$c^2$    &
               &
               &
\vspace*{0mm} \\
            Number of events                 &
           $1581$                            &
           $1926$                            &
           $ 689$    
\vspace*{0mm} \\
            Signal fraction                  &
           $0.902\pm0.015$                   &
           $0.937\pm0.007$                   &
           $0.917\pm0.010$
\vspace*{1mm} \\
\hline \hline
\end{tabular}
\end{table}

\begin{figure}[!t]
  \centering
\hspace*{-1mm}
  \includegraphics[width=0.32\textwidth]{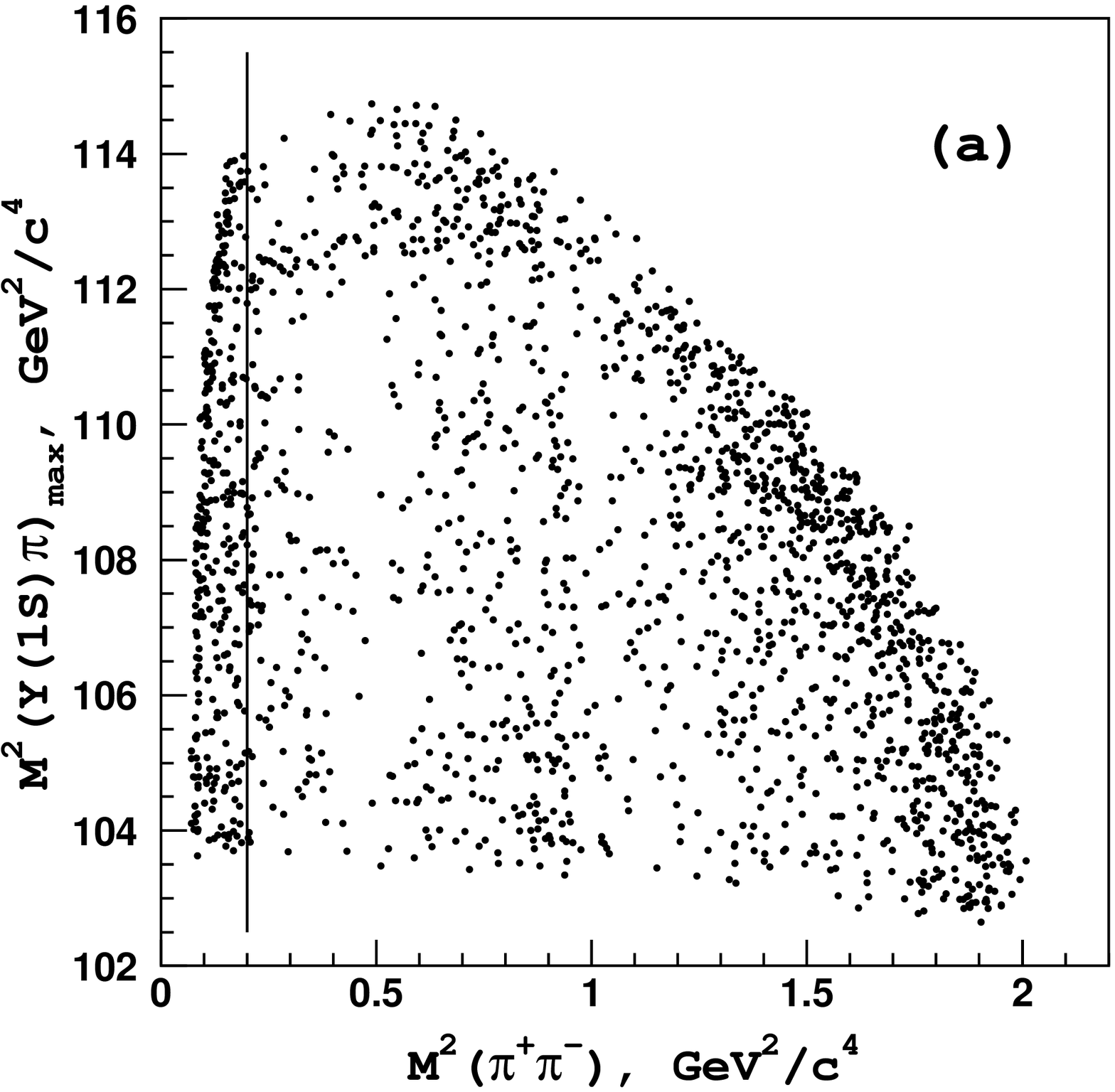} \hfill
  \includegraphics[width=0.32\textwidth]{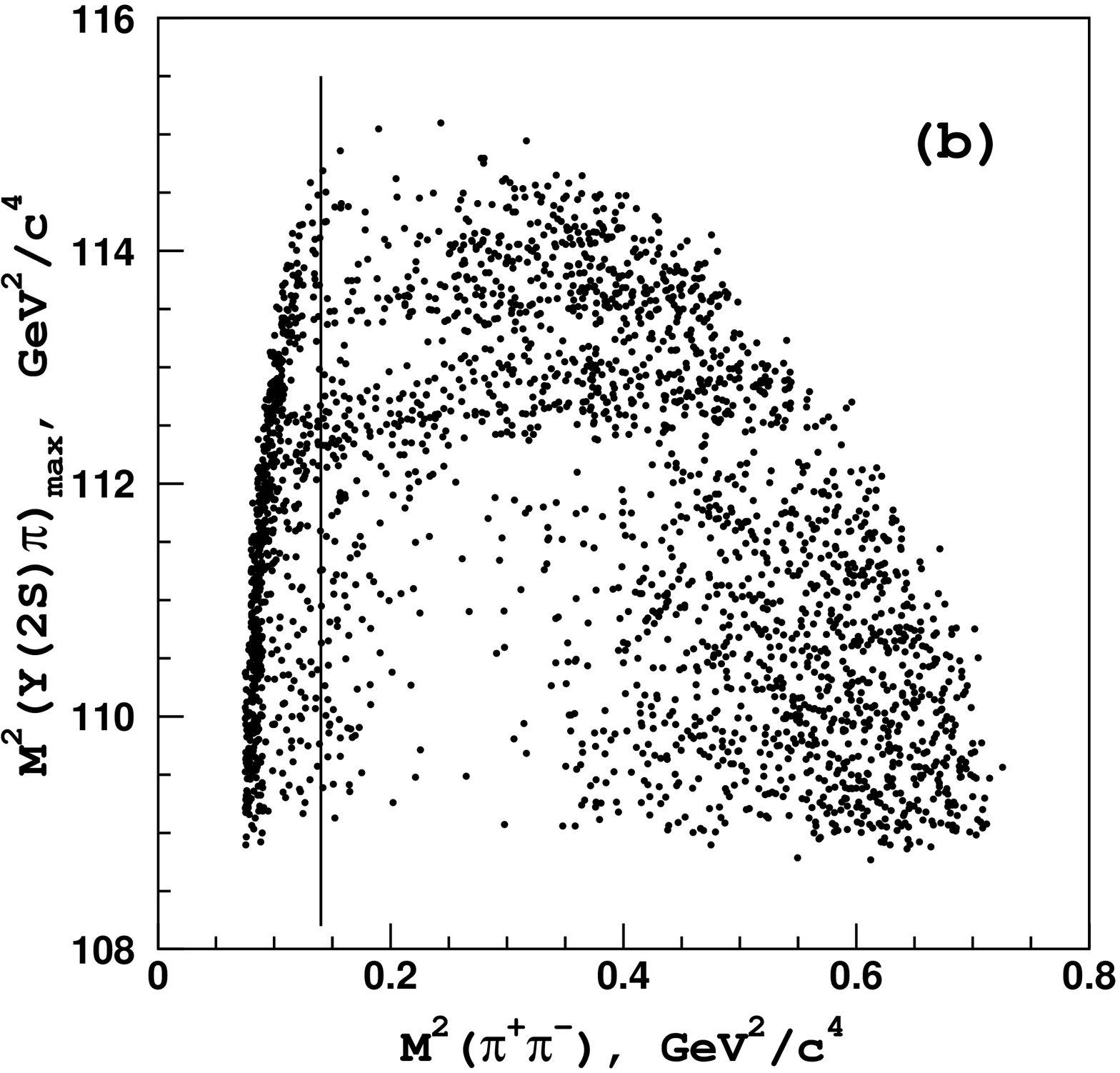} \hfill
  \includegraphics[width=0.32\textwidth]{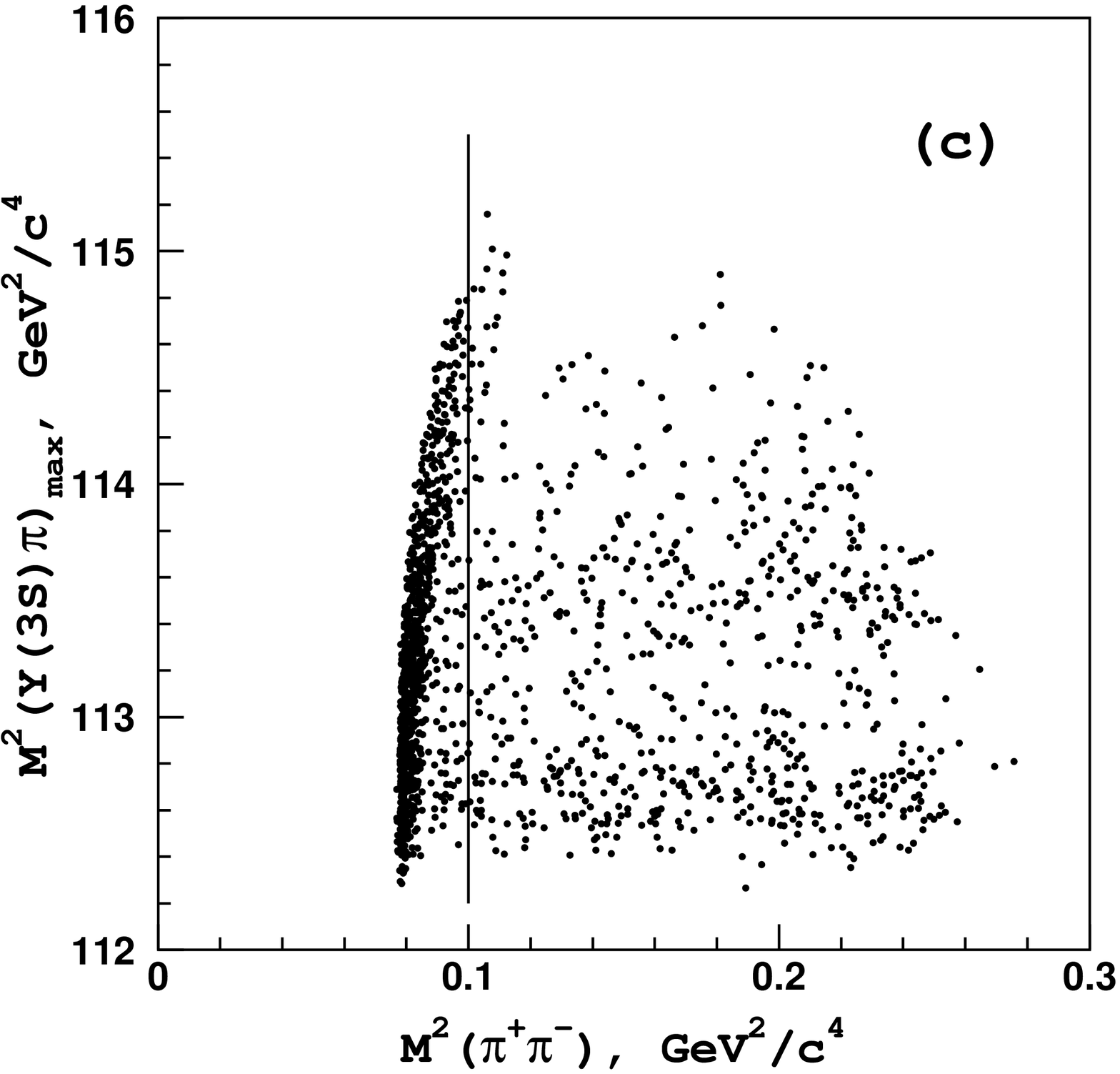}
  \caption{Dalitz plots for $\Un\pp$ events in the (a) $\Uo$; (b)
    $\Ut$; \mbox{(c) $\Uth$} signal regions. Dalitz plot regions to
    the right of the vertical lines are included in the amplitude
    analysis.}
\label{fig:ynspp-s-dp}
\end{figure}

Figure~\ref{fig:ynspp-s-dp} shows Dalitz plots of the events in the signal
regions for the three decay channels under study. In all cases, two 
horizontal bands are evident in the $\Un\pi$ system near $10.61\,\gevm$ 
($\sim112.6$~GeV$^2/c^4$) and $10.65\,\gevm$ ($\sim113.3$~GeV$^2/c^4$). 
In the following we refer to these structures as $Z_b(10610)$ and 
$Z_b(10650)$, respectively.

\begin{figure}[!t]
  \centering
  \includegraphics[width=0.32\textwidth]{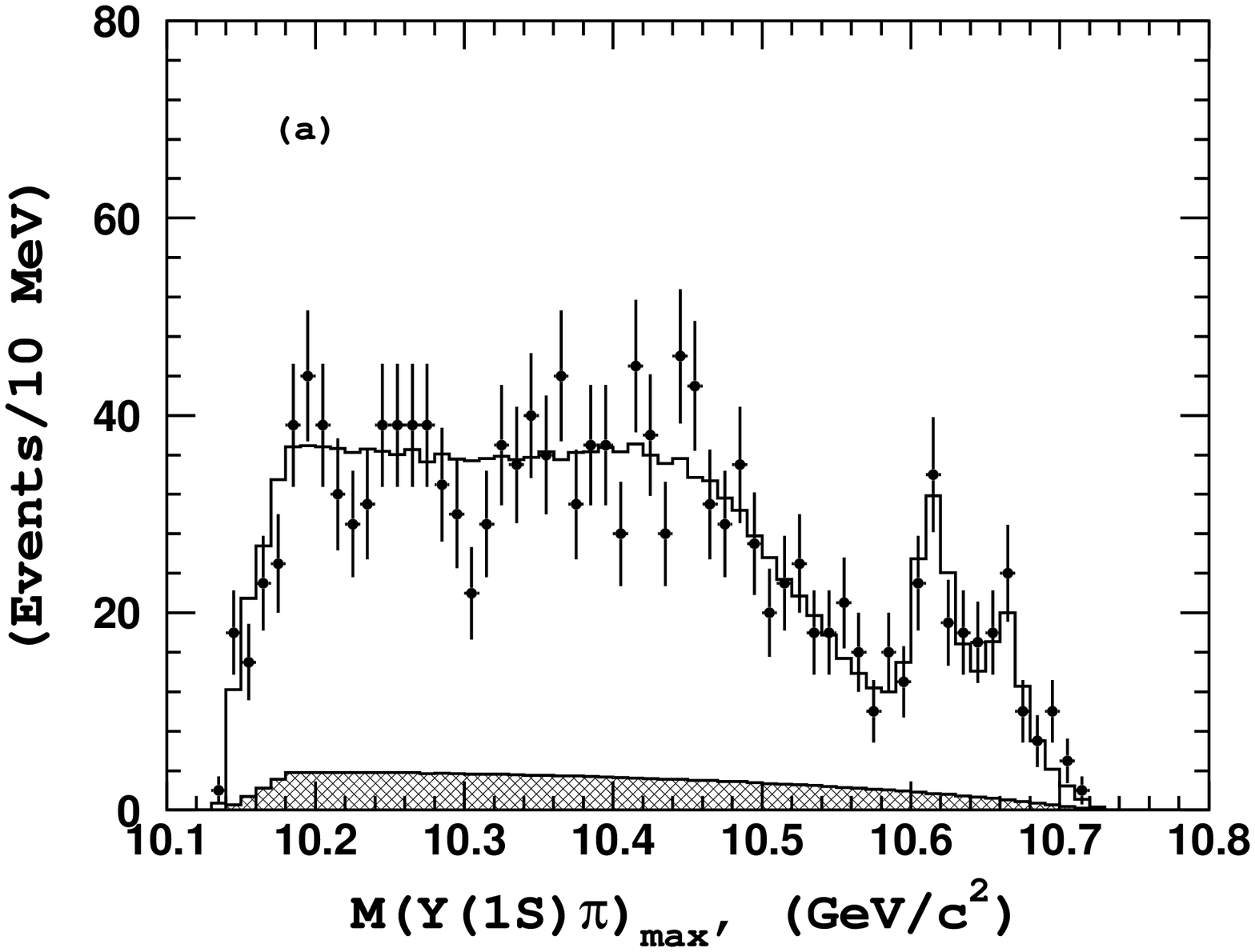} \hfill
  \includegraphics[width=0.32\textwidth]{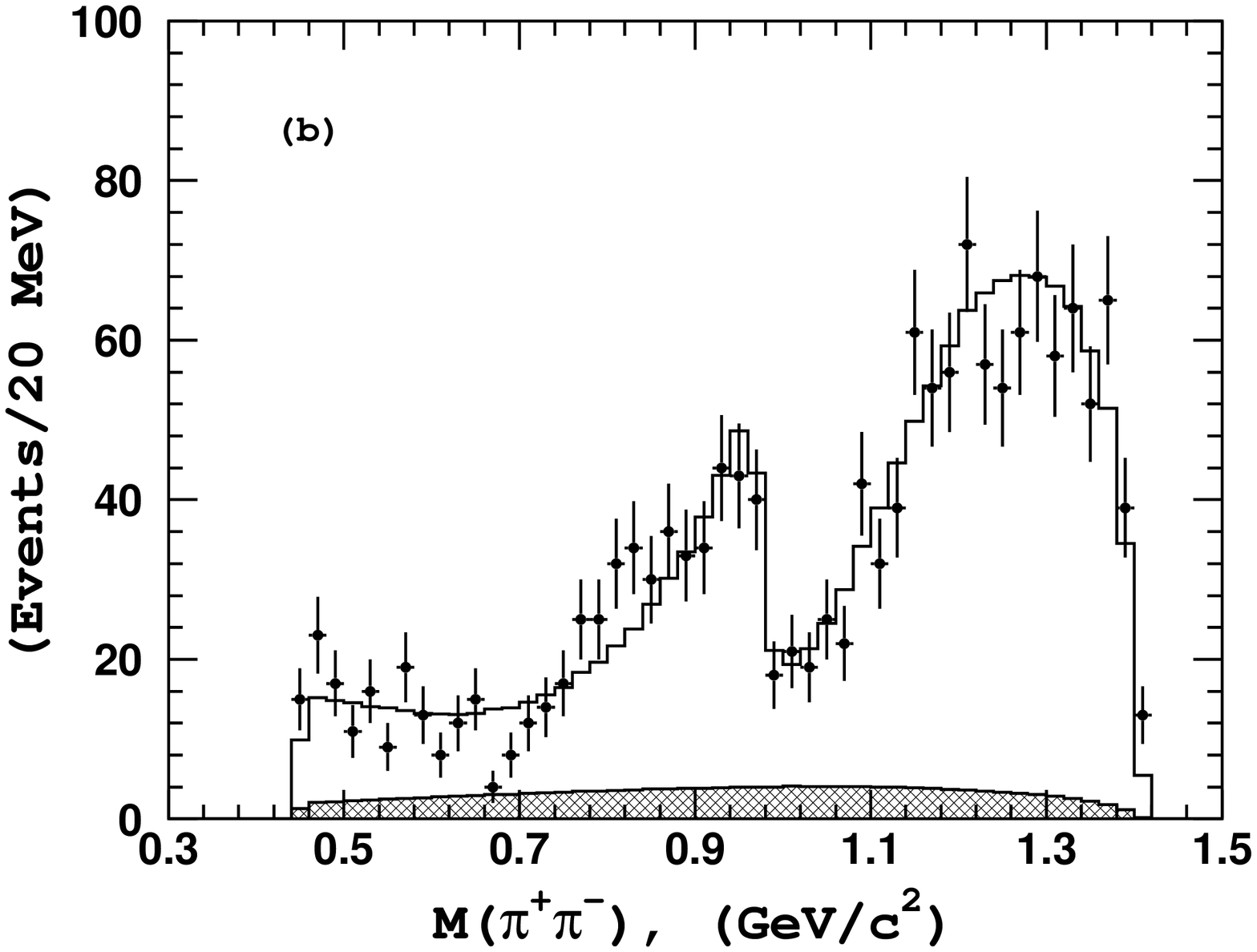} \hfill
  \includegraphics[width=0.32\textwidth]{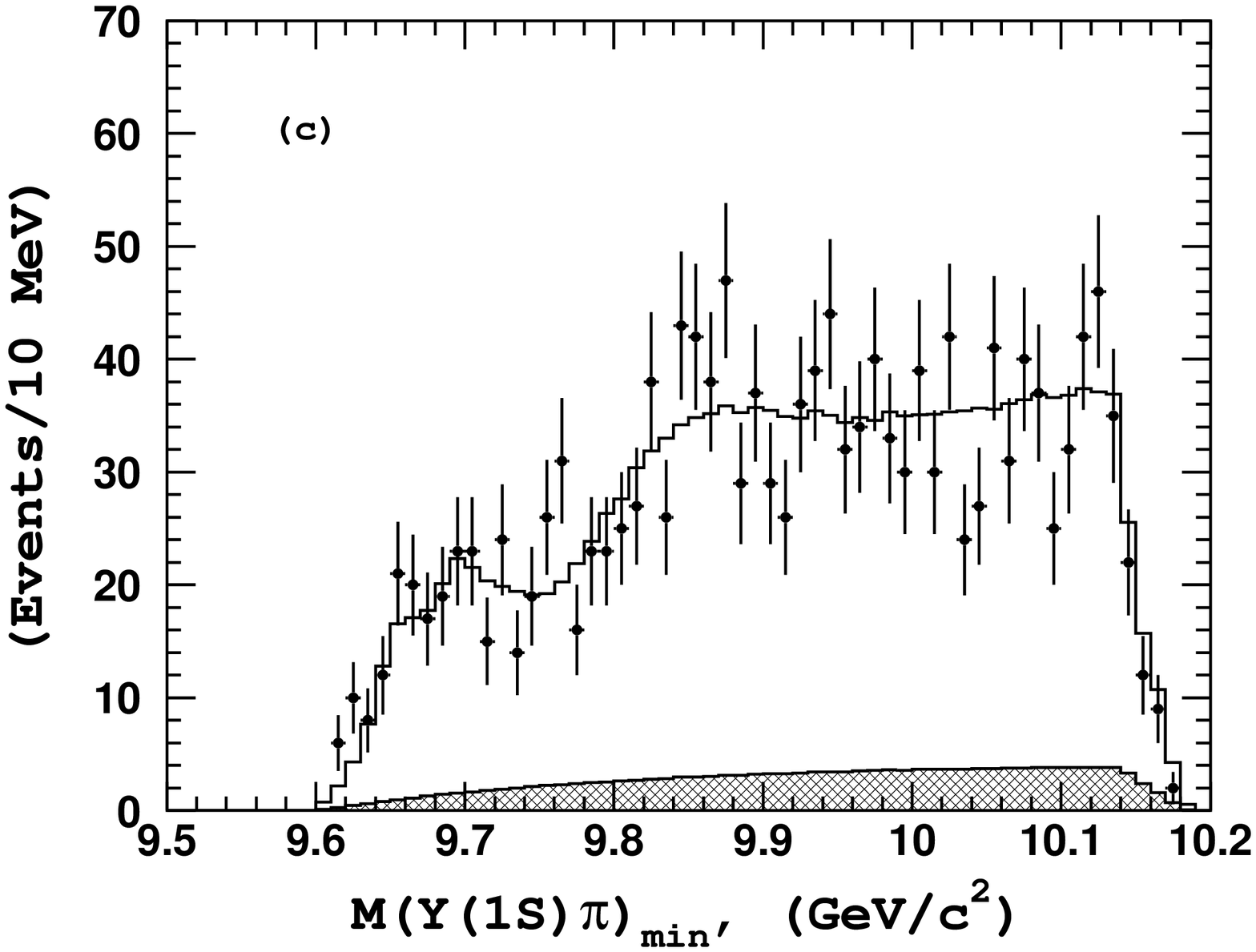} \\
  \includegraphics[width=0.32\textwidth]{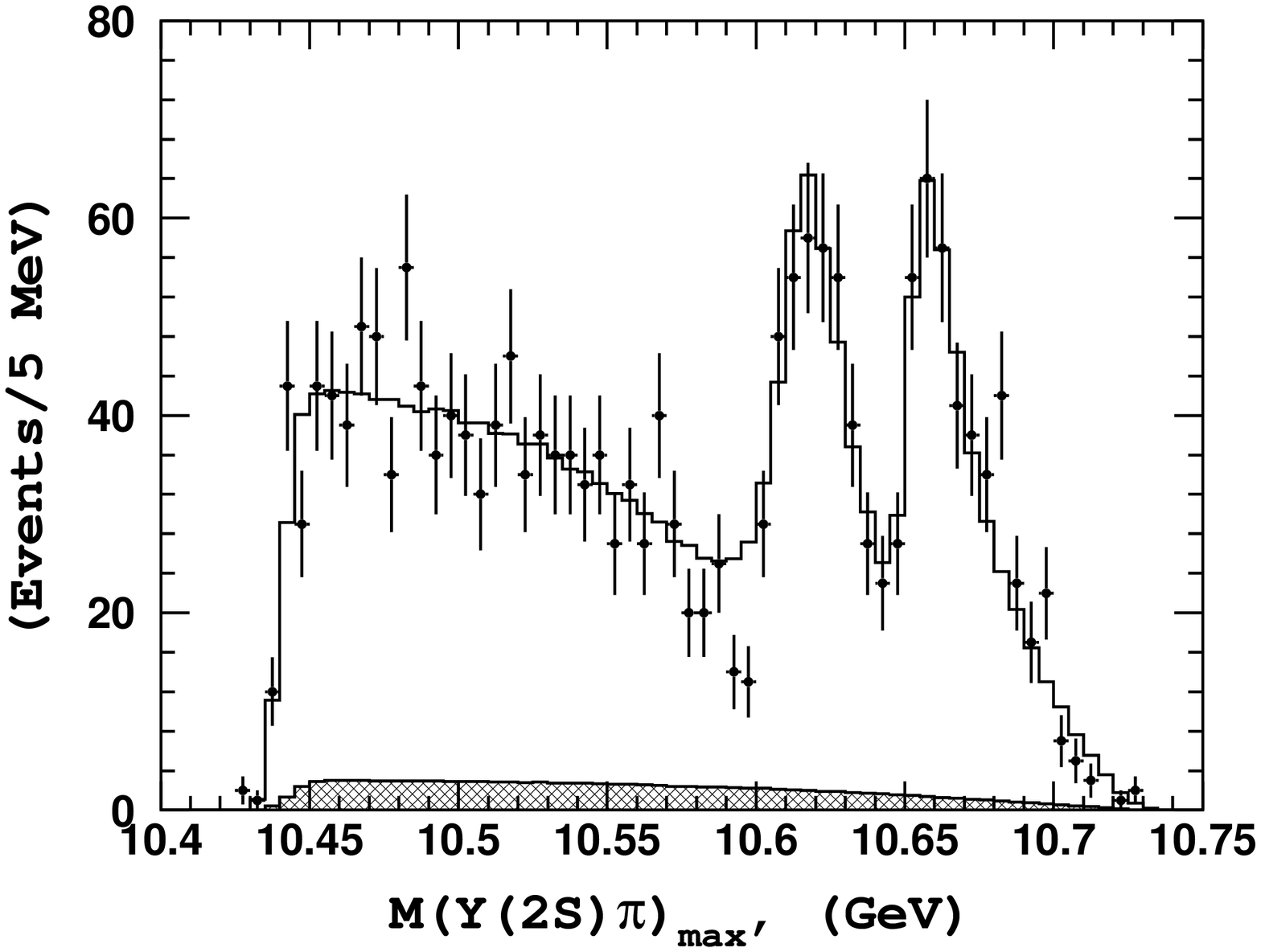} \hfill
  \includegraphics[width=0.32\textwidth]{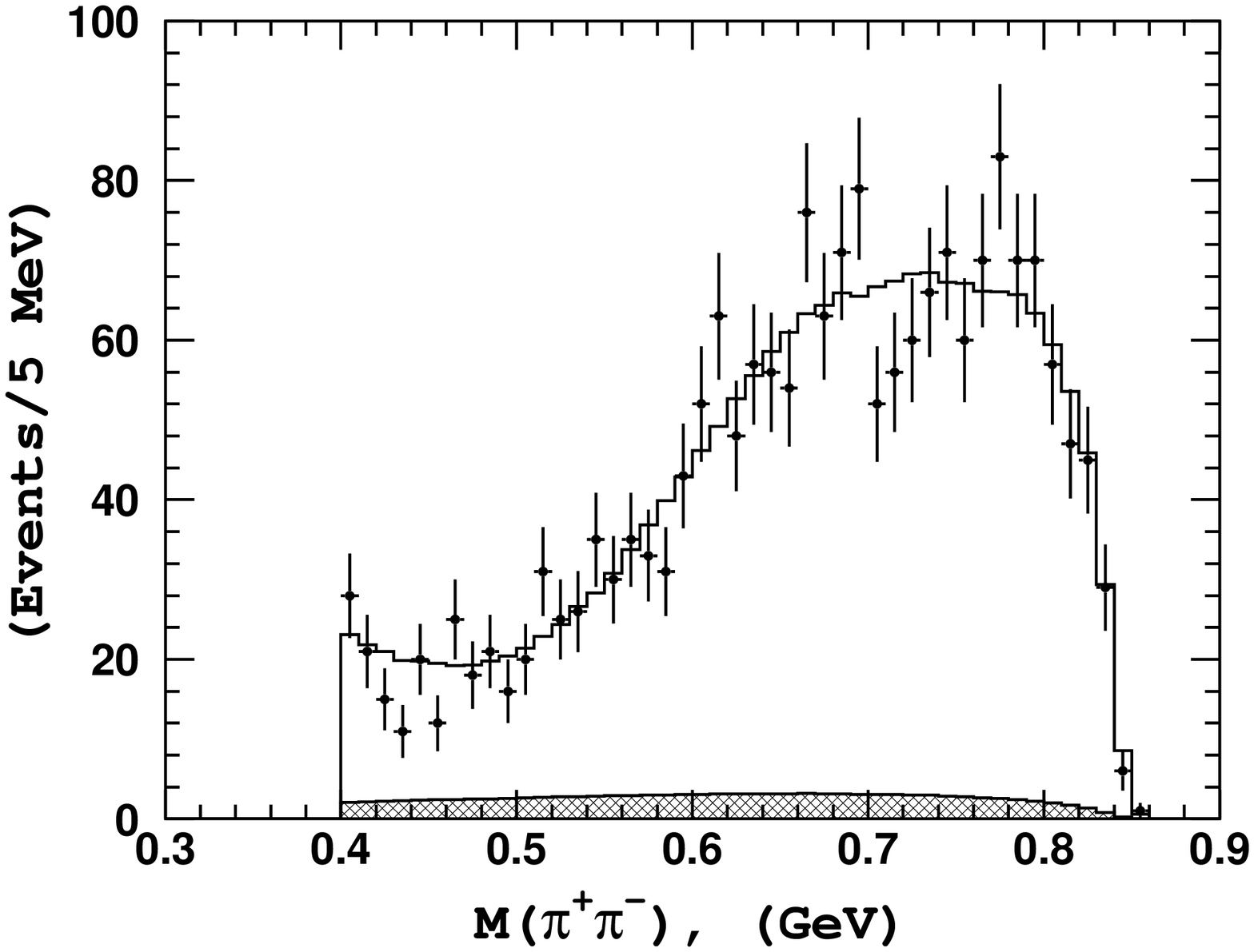} \hfill
  \includegraphics[width=0.32\textwidth]{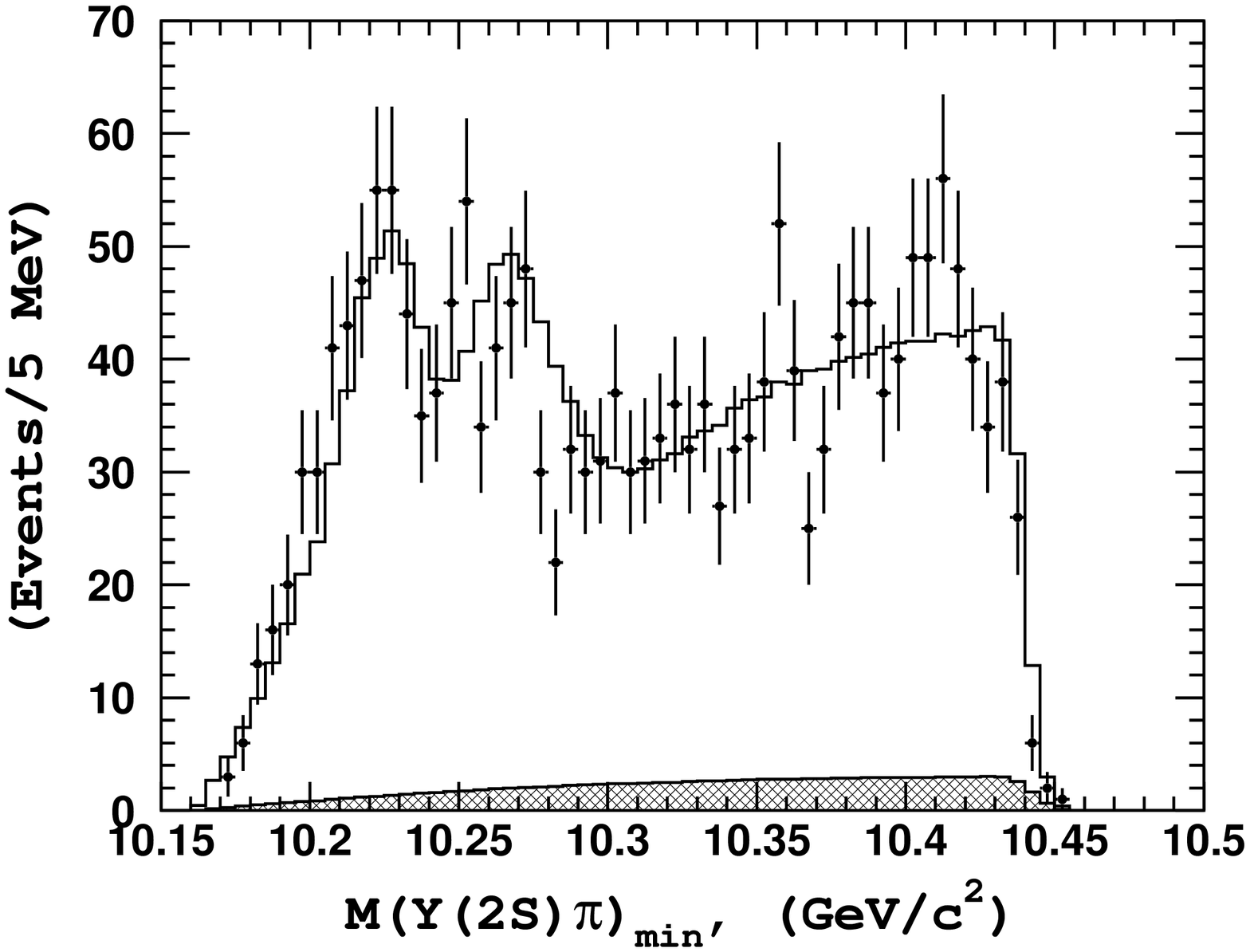} \\
  \includegraphics[width=0.32\textwidth]{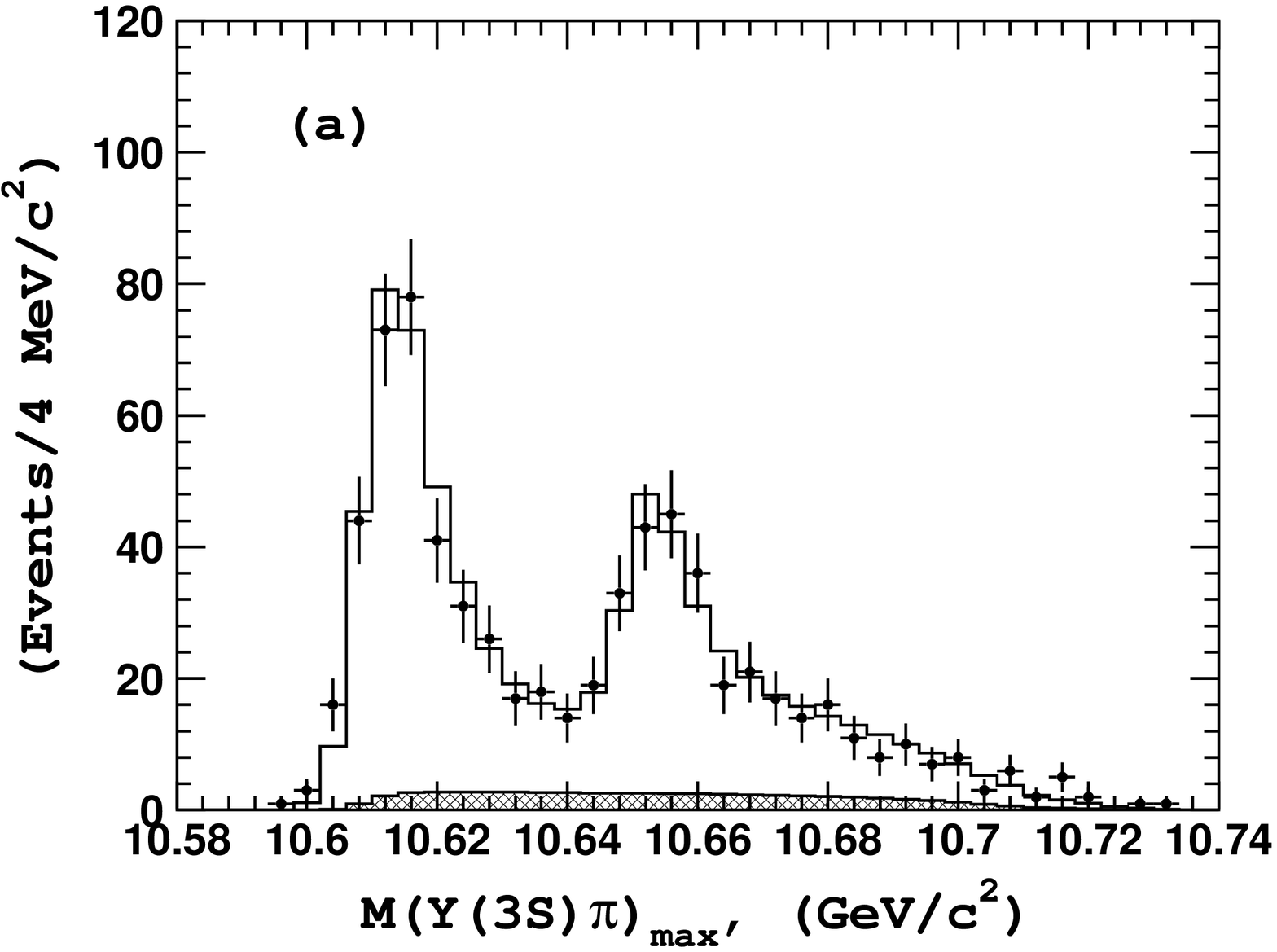} \hfill
  \includegraphics[width=0.32\textwidth]{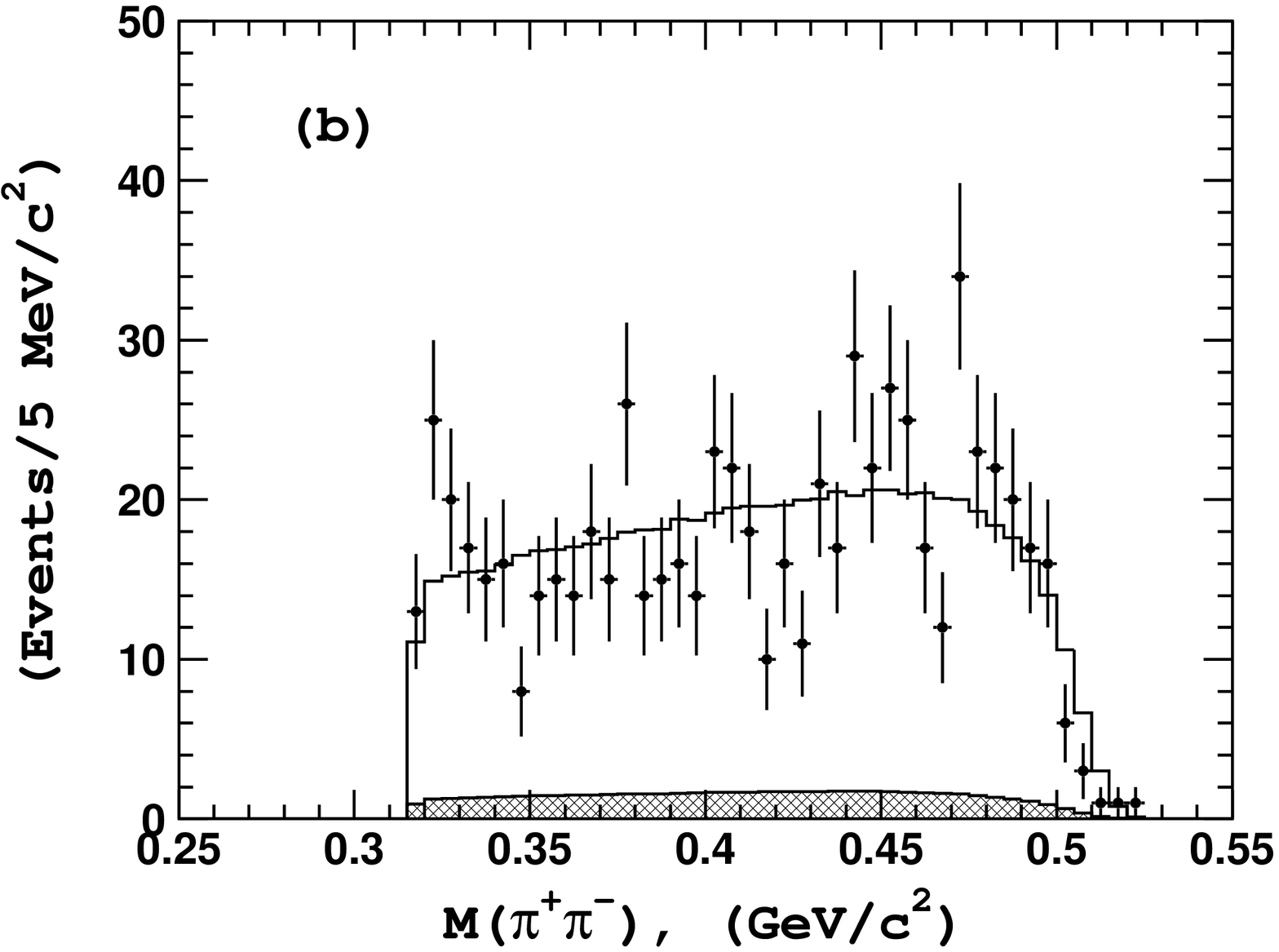} \hfill
  \includegraphics[width=0.32\textwidth]{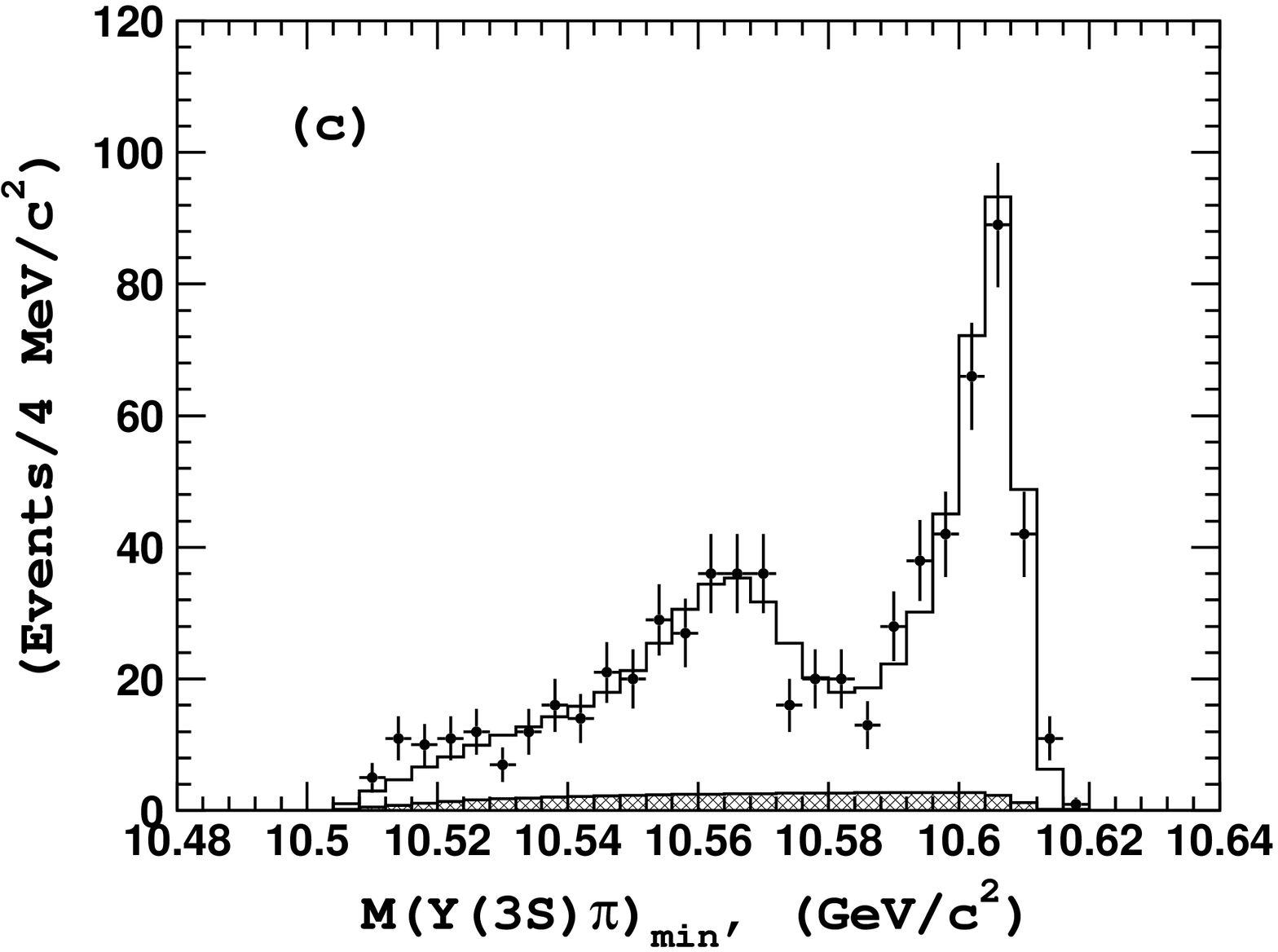}
  \caption{Comparison of fit results (open histogram) with
    experimental data (points with error bars) for events in the $\Uo$
    (top row), $\Ut$ (middle row), and $\Uth$ (bottom top) signal
    regions.  The hatched histogram show the background component.}
\label{fig:y3spp-f-hh}
\end{figure}

We use the following parameterization for the $\Uf\to\Un\pp$ three-body
decay amplitude:
\begin{equation}
M(s_1,s_2) = A_1(s_1,s_2) + A_2(s_1,s_2) + A_{f_0} + A_{f_2} + A_{NR},
\label{eq:model}
\end{equation}
where $s_1 = M^2(\Un\pi^+)$, $s_2 = M^2(\Un\pi^-)$. Here we assume that
the dominant contributions come from amplitudes that conserve the 
orientation of the spin of the heavy quarkonium state and, thus, both pions
in the cascade decay $\Uf\to Z_b\pi\to\Un\pp$ are emitted in an $S$-wave
with respect to the heavy quarkonium system. As will be shown later,
angular analyses support this assumption. Consequently, we parameterize
the observed $Z_b(10610)$ and $Z_b(10650)$ peaks with an $S$-wave 
Breit-Wigner function
\begin{equation}
BW(s,M,\Gamma)=\frac{\Gamma}{M^2-s-iM\Gamma},
\end{equation}
where we do not consider possible $s-$dependence of the resonance
width $\Gamma$. To account for the possibility of $\Uf$ decay to both
$Z^+\pi^-$ and $Z^-\pi^+$, the amplitudes $A_1$ and $A_2$ are
symmetrized with respect to $\pi^+$ and $\pi^-$ transposition. Taking
into account isospin symmetry, the resulting amplitude is written as
\begin{equation}
A_k(s_1,s_2) = a_k e^{i\delta_k} (BW(s_1,M_k,\Gamma_k) + BW(s_2,M_k,\Gamma_k)),
\end{equation}
where the masses $M_k$ and the widths $\Gamma_k$ ($k = 1,2$) are free 
parameters of the fit. Due to the very limited phase space available in
the $\Uf\to\Uth\pp$ decay, there is a significant overlap between the two
processes $\Uf\to Z^+_{b}\pi^-$ and $\Uf\to Z^-_{b}\pi^+$. We also include
$A_{f_0}$ and $A_{f_2}$ amplitudes to account for possible contributions
in the $\pp$ channel from $f_0(980)$ scalar and $f_2(1270)$ tensor states,
respectively. Inclusion of the $f_0(980)$ state is necessary in order to
describe the prominent structure in the $M(\pp)$ spectrum for the $\Uo\pp$
final state around $M(\pp)=1.0\,\gevm$ (see Fig.~\ref{fig:y3spp-f-hh}). 
We also find that the addition of the $f_2(1270)$ gives a better description
of the data at $M(\pp)>1.0\,\gevm$ and drastically improves the fit
likelihood values. We use a Breit-Wigner function to parameterize the
$f_2(1270)$ and a coupled-channel Breit-Wigner (Flatte) function~\cite{Flatte}
for the $f_0(980)$. The mass and the width of the $f_2(1270)$ state are
fixed at their world average values~\cite{PDG}; the mass and the coupling
constants of the $f_0(980)$ state are fixed at values defined from the
analysis of  $B^+\to K^+\pp$: $M(f_0(980))=950$~MeV/$c^2$, $g_{\pi\pi}=0.23$,
$g_{KK}=0.73$~\cite{kpp}.

Following suggestions given in Refs.\cite{Voloshin:2007dx}
and references therein, the non-resonant amplitude $A_{NR}$ has been 
parameterized as 
\begin{equation}
A_{\rm NR} = a^{\rm nr}_1\, e^{i\delta^{\rm nr}_1} +
             a^{\rm nr}_2\, e^{i\delta^{\rm nr}_2} \, s_3,
\end{equation}
where $s_3 = M^2(\pp)$ ($s_3$ is not an independent variable and can be
expressed via $s_1$ and $s_2$ but we use it here for clarity),
$a^{\rm nr}_1$, $a^{\rm nr}_2$, $\delta^{\rm nr}_1$ and $\delta^{\rm nr}_2$
are free parameters of the fit (with an exceptions of the $\Uth\pp$ channel
as described below).

The logarithmic likelihood function ${\cal{L}}$ is then constructed as 
\begin{equation}
{\cal{L}} = -2\sum{\log(f_{\rm sig}S(s_1,s_2) + (1-f_{\rm sig})B(s_1,s_2))},
\end{equation}
where $S(s_1,s_2) = |M(s_1,s_2)|^2$ folded with the detector resolution 
function (2~MeV/$c^2$ for $M(\Un\pi^\pm)$ when both $\Un$ and $\Uf$ mass 
constraints are used; the $M(\pp)$ resolution is
not taken into account since no narrow resonances are observed
in the $\pp$ system), $B(s_1,s_2)=1$ and $f_{\rm sig}$ is a fraction of
signal events in the data sample. Both $S(s_1,s_2)$ and $B(s_1,s_2)$ are
corrected for reconstruction efficiency.

In the fit to the $\Uo\pp$ sample, the amplitudes and phases of all of
the components are allowed to float. However, in the $\Ut\pp$
and $\Uth\pp$ samples the available phase space is significantly smaller and
contributions from the $f_0(980)$ and $f_2(1270)$ are not well
defined.  Thus, in the fit to the $\Ut\pp$ and $\Uth\pp$ signal
samples, we fix the amplitudes and relative phases of these components
to the values measured in the fit to the $\Uo\pp$ sample. Moreover, in
the fit to the $\Uth\pp$ sample, we also fix the $a_2^{\rm nr}$ and
$\delta_2^{\rm nr}$ parameters of the $A_{\rm nr}$ amplitude. Possible
effects of these assumptions are considered while determining the
model-dependent uncertainty. Results of the fits to $\Uf\to\Un\pp$
signal events are shown in Fig.~\ref{fig:y3spp-f-hh}, where
one-dimensional projections of the data and fits are compared.  To
combine $Z^+_b$ and $Z^-_b$ events we plot $\Un\pi$ mass distributions
in terms of $M(\Un\pi)_{\min}$ and $M(\Un\pi)_{\max}$; fits are
performed in terms of $M(\Un\pi^+)$ and $M(\Un\pi^-)$. Results of the
fits are summarized in Table~\ref{tab:results}. We try various
alternative models to parameterize the decay amplitude as described in
the systematic uncertainty section. The combined statistical
significance of the two peaks exceeds $10\,\sigma$ for all tested models
and for all $\Un\pp$ channels.

The main sources of systematic uncertainties are found to be:
\begin{itemize}
  \item{Dependence of reconstruction efficiency on the Dalitz plot variables
has been determined from MC simulation of signal events uniformly distributed
over the Dalitz plot. Since the pions from the $\Uf\to\Un\pp$ decays have low
momenta, the efficiency determined from MC might be systematically shifted
by up to 5-10\%. The effect on the parameters of the observed peaks is estimated by
varying the efficiency at the edges of the Dalitz plot by $\pm10$\% and 
refitting the data.}
  \item{To estimate the effect of the parameterization used to fit the data
on the parameters of the $Z_b$ peaks we fit the data with 
modifications of the nominal model described in Eq.~\ref{eq:model}. In 
particular, we vary the $M(\pp)$ dependence of the non-resonant amplitude 
$A_{\rm NR}$, try to include a $D$-wave component into $A_{\rm NR}$, etc.
The variations in the extracted $Z_b$ parameters determined from fits with
modified models are taken as estimates of the model uncertainties.}
  \item{The uncertainty in the c.m.\ energy leads to an uncertainty in 
Dalitz plot boundaries. This effect is particularly important for the 
$\Uth\pp$ channel. To estimate the associated effect on the $Z_b$ parameters,
we generate normalization phase space MC samples that correspond to 
$E_{c.m.}\pm 3$~MeV, where $E_{c.m.}$ is the nominal c.m.\ energy.}
\end{itemize}

Systematic effects associated with uncertainties in the description of 
the combinatorial background is found to be negligible. The results of 
our studies on systematic uncertainties are summarized in 
Table~\ref{tab:ynspp_syst}.

\begin{table}[!t]
  \caption{Summary of dominant sources of systematic uncertainties.
Uncertainties for masses and widths are given in $\mevm$ and in degrees
for the relative phase.
Quoted numbers are for $\Uo/\Ut/\Uth$ channels, respectively.} 
  \medskip
  \label{tab:ynspp_syst}
\centering
  \begin{tabular}{lcccc} \hline \hline
              Parameter                                           &
              Efficiency                                          &
              Model                                               &
              $E_{\rm c.m.}$                                      &
              Total                 
\\ \hline
         $M(Z_b(10610))$                                          &
         $\pm 1$/$\pm 1$/$\pm 1$                                  &
         $\pm 1$/$^{+2}_{-3}$/$^{+4}_{-0}$                        &
         $\pm 1$/$\pm1$/$\pm 2$                                   &
         $\pm 2$/$^{+3}_{-4}$/$^{+5}_{-2}$            

\vspace*{-1mm} \\
         $\Gamma(Z_b(10610))$                                     &
         $\pm 1$/$\pm 1$/$\pm 1$                                  &
         $\pm 2$/$^{+0.3}_{-1.4}$/$^{+2.4}_{1.2}$                 &
         $\pm 1$/$\pm2$/$\pm 3$                                   &
         $\pm 2$/$^{+2}_{-3}$/$\pm 4$                 

\vspace*{-1mm} \\
         $M(Z_b(10650))$                                          &
         $\pm 1$/$\pm 1$/$\pm 1$                                  &
         $\pm 1$/$^{+1}_{-1}$/$^{+2}_{-1}$                        &
         $\pm 1$/$\pm1$/$\pm 1$                                   &
         $\pm 2$/$\pm 2$/$\pm 2$                      
\vspace*{1mm} \\
         $\Gamma(Z_b(10650))$                                     &
         $\pm 1$/$\pm 1$/$\pm 1$                                  &
         $\pm 2$/$^{+1.4}_{-4.6}$/$^{+3.7}_{-0.2}$                &
         $\pm 1$/$\pm3$/$\pm 1$                                   &
         $\pm 3$/$^{+4}_{-6}$/$^{+4}_{-2}$            
\vspace*{1mm} \\
           Rel.  phase                                            &
         $\pm 3$/$^{+6}_{-4}$/$\pm 5$                             &
         $^{+0}_{-50} $/$^{+11}_{-6}$/$^{+21}_{-58}$              &
         $\pm 3$/$\pm6$/$\pm 9$                                   &
         $^{+5}_{-50}$/$^{+14}_{-9}$/$^{+23}_{-59}$   
\vspace*{1mm} \\
           Rel.  amp.                                             &
         $\pm0.01/\pm0.02/\pm0.02$                                &
         $^{+0.08}_{-0.0}/^{+0.03}_{-0.0}/^{+0.14}_{-0.02}$       &
         $\pm 0.02/\pm0.02/\pm0.04$                               &
         $^{+0.09}_{-0.03}/^{+0.04}_{-0.03}/^{+0.15}_{-0.0.05}$
\\
\hline \hline
\end{tabular}
\end{table}

\section{\boldmath Analysis of $\Upsilon(5S)\to h_b(1P,2P)\pp$}

$\hb$ and $\hbp$ have been recently observed by Belle in the decay
$\Uf\to\hbn\pp$~\cite{Belle_hb}. Here we study resonant structure of
these decays. Because of high background Dalitz plot analysis is
impossible with current statistics, therefore we study the one
dimensional distributions in $M(\hbn\pi)$. We use the same selection
requirements and reconstruction procedure as in Ref.~\cite{Belle_hb}.

The $\Uf\to\hbn\pp$ decays are reconstructed inclusively using the
missing mass of the $\pp$ pair. We select $\pip$ and $\pim$ candidates
that originate from the vicinity of the interaction point ($dr<3\mm$,
$|dz|<2\cm$) and are positively identified as pions. We reject tracks
that are identified as electrons.  The continuum $\ee\to q\bar{q}$
($q=u,\;d,\;s,\;c$) background is suppressed by a requirement on the
ratio of the second to zeroth Fox-Wolfram moments
$R_2<0.3$~\cite{Fox-Wolfram}.

\begin{figure}[!tbp]
\includegraphics[width=0.32\textwidth]{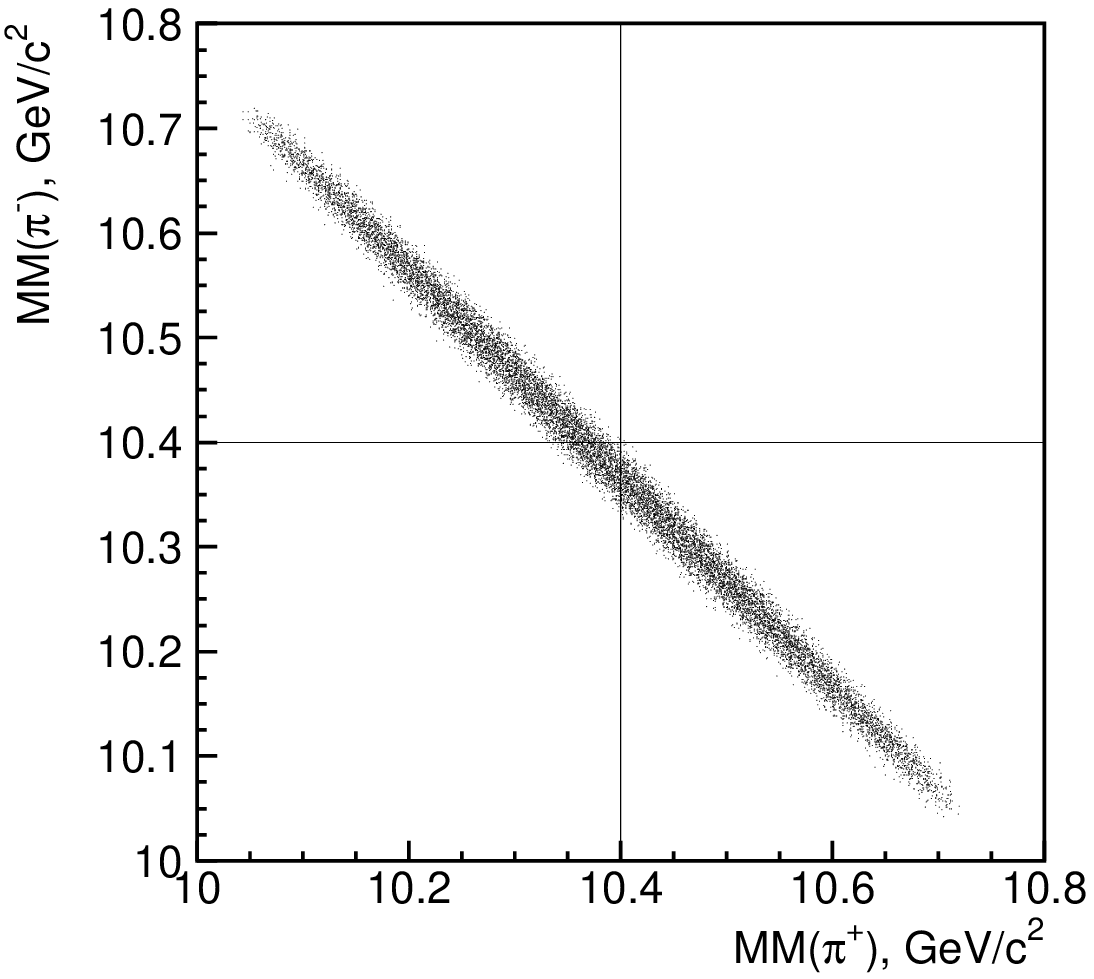} \hfill
\includegraphics[width=0.32\textwidth]{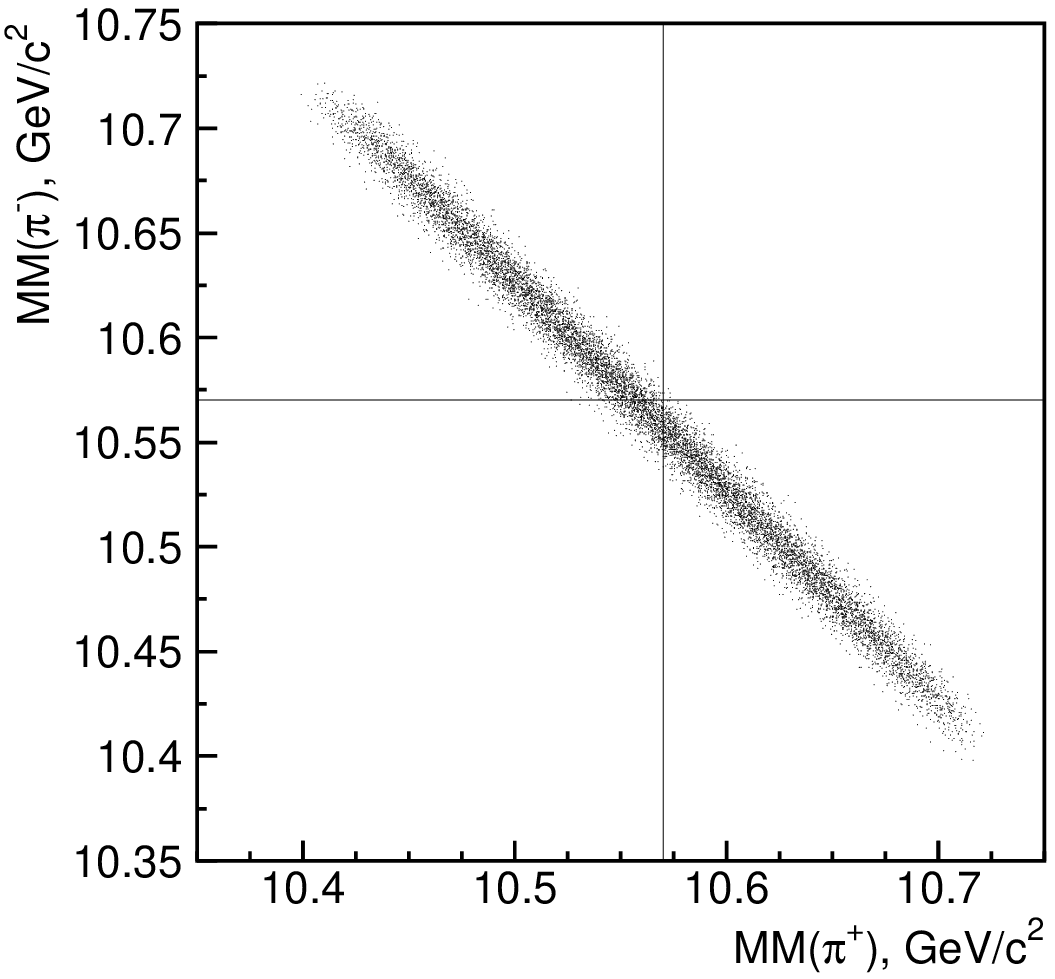} \hfill
\includegraphics[width=0.32\textwidth]{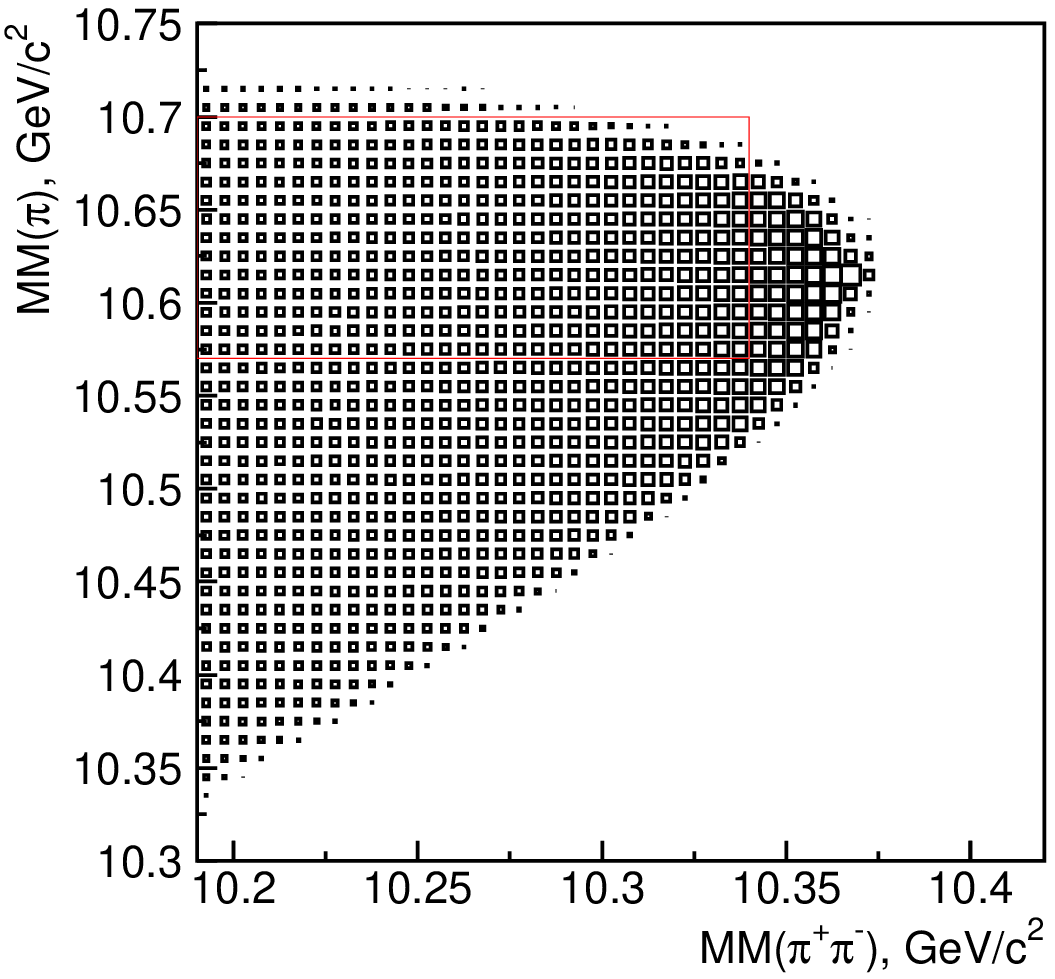}
\caption{The $\mmpim$ versus $\mmpip$ distribution for the
  $\Uf\to\hb\pp$ (left) and $\Uf\to\hbp\pp$ (middle) events simulated
  according to the phase-space MC model. Lines indicate the values of
  cuts used to avoid double counting after symmetrization relative to
  $\pip$ and $\pim$. Right: the $\mmp$ versus $\mmpp$ distribution for
  inclusive $\ks\to\pp$ signal in the fast MC simulation. The red
  rectangle shows the fit region for the $\hbp\pp$ analysis. }
\label{mc_hb_signal}
\end{figure}

We define the $M(\hbn\pip)$ as a missing mass of the opposite sign
pion, $MM(\pim)$.  We measure the yield of signal decays as a function
of the $MM(\pipm)$ by fitting the $\mmpp$ spectra in the bins of
$MM(\pipm)$. We combine the $\mmpp$ spectra for the corresponding
$MM(\pip)$ and $MM(\pim)$ bins and we use half of the phase-space to
avoid double counting, as shown in Fig.~\ref{mc_hb_signal}.

In the studies of the $\Uf\to\hb\pp$ channel we subdivide the interval
$10.4\,\mevm<\mmp<10.72\,\mevm$ into 32 bins and perform a $\chi^2$
fit to the $\mmpp$ spectrum for each $\mmp$ bin. The fit function
consists of four components: the $\hb$ signal, the $\Ut$ signal, a
reflection from the $\Uth\to\Uo\pp$ decay, and combinatorial
background. The shapes of the signal components and $\Uth$ reflection
is determined from the analysis of exclusive $\uu\pp$ data as
described in Ref.~\cite{Belle_hb}.  The signals are parameterized by a
Crystal Ball function to accommodate tails due to initial state
radiation of soft photons that accounts for about 8\% of the signal
yield. The resolution width of the $\Ut$ ($\hb$) is
$\sigma=6.5\,\mevm$ ($6.8\,\mevm$). The $\Uth\to\Uo\pp$ reflection is
described by a single Gaussian function with the width of
$\sigma=18\,\mev$.  The ratio of the $\Uth\to\Uo\pp$ and $\Ut$ yields
is measured from $\uu\pp$ data and corrected for the branching
fractions for $\Uo\to\uu$ and $\Ut\to\uu$ decays. In the fit to the
inclusive $MM(\pp)$ spectra, this ratio is allowed to float within the
uncertainties of the exclusive $\uu\pp$ measurements. The
combinatorial background is parameterized by a Chebyshev
polynomial. The polynomial order decreases monotonically from 
10-th order for the first bin to 6-th order for the 11-th bin. The
background shape in the first $\mmp$ bins is complicated due to the
proximity of the kinematic boundary, therefore a high-order polynomial
is used.

\begin{figure}[!tbp]
\includegraphics[width=0.32\textwidth]{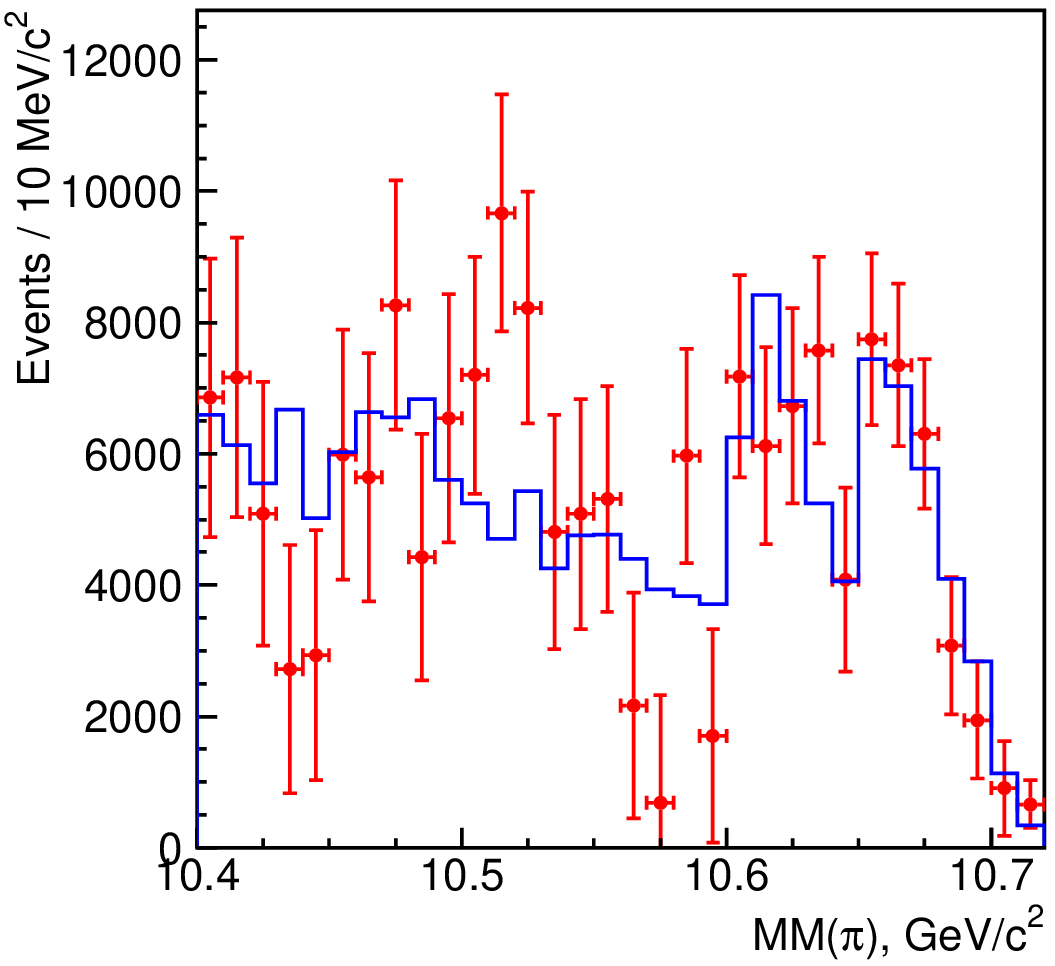} \hfill
\includegraphics[width=0.32\textwidth]{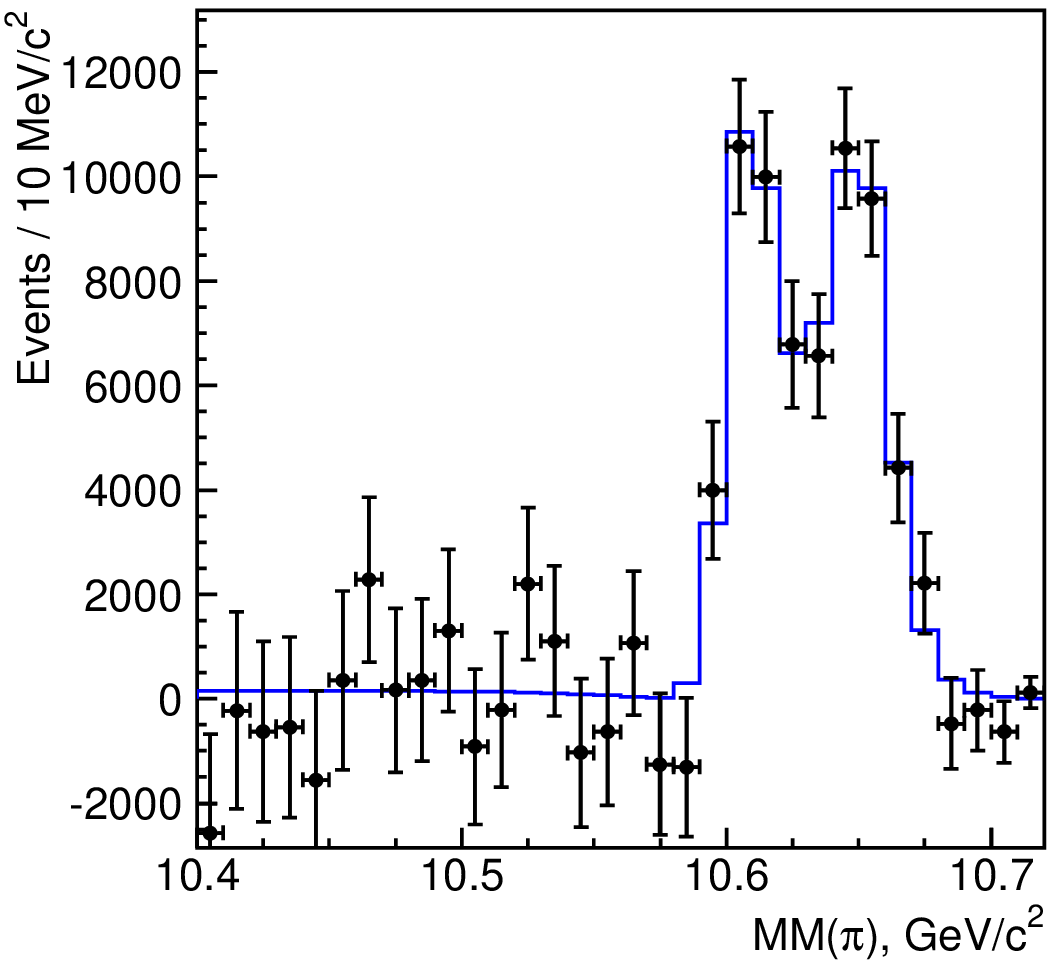} \hfill
\includegraphics[width=0.32\textwidth]{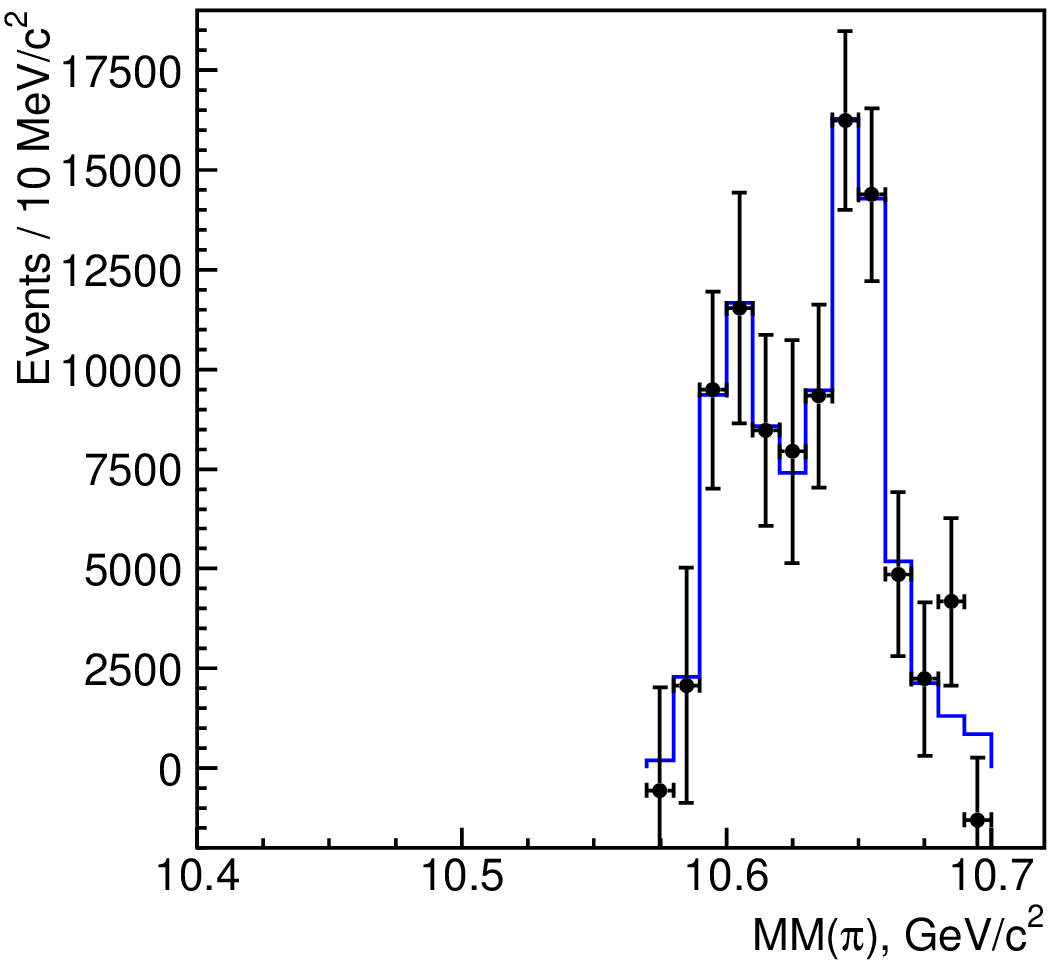}
\caption{Left: the yield of $\Ut$ as a function of the $\mmp$ measured
  using the inclusive data (points with error bars) and exclusive
  $\uu\pp$ data (histogram).  Middle: the yield of the $\hb$ as a
  function of $\mmp$ (points with error bars) and results of the fit
  (histogram).  Right: the yield of the $\hbp$ as a function of $\mmp$
  (points with error bars) and results of the fit (histogram).}
\label{nhb_n2s_vs_mmp}
\end{figure}

Results of fits for the $\hb$ and $\Ut$ yield as a function of
$MM(\pi)$ are shown in Fig.~\ref{nhb_n2s_vs_mmp}, where the $\Ut$
yield measured from the analysis of the exclusive $\Uf\to\uu\pp$ data
and renormalized to the total number of events in inclusive spectrum
is also shown for comparison. The $\Ut$ yields for inclusive and
exclusive measurements agree well. The $\hb$ yield shown in
Fig.~\ref{nhb_n2s_vs_mmp} exhibits a clear two-peak structure without
any significant non-resonant contribution. The peak positions are
consistent with those for the $\zbo$ and $\zbt$ observed in the
$\Un\pipm$ final states.

We perform a $\chi^2$ fit to the $\mmp$ distribution.  We assume that
spin-parity for both $\zbo$ and $\zbt$ is $J^P=1^+$, therefore in the
fit function we use a coherent sum of two $P$-wave Breit-Wigner
amplitudes; we add also a non-resonant contribution.
\begin{equation}
f=A\,|BW(s,M_1,\Gamma_1)+ae^{i\phi}BW(s,M_2,\Gamma_2)
+be^{i\psi}|^2\;\frac{qp}{\sqrt{s}}.
\end{equation}
Here $\sqrt{s}\equiv\mmp$; the variables $A$, $M_k$, $\Gamma_k$
($k=1,2$), $a$, $\phi$, $b$ and $\psi$ are floating in the fit;
$\frac{qp}{\sqrt{s}}$ is a phase-space factor, $p$ ($q$) is the
momentum of the pion originating from the $\Uf$ ($\zb$) decay measured
in the rest frame of the corresponding mother particle.

The $P$-wave Breit-Wigner amplitude is expressed as
\begin{equation}
BW(s,M,\Gamma)=\frac{q/\sqrt{s}\,F}{M^2-s-iM\,\Gamma(s)}.
\end{equation}
Here $F$ is the $P$-wave Blatt-Weisskopf form-factor
$F=\sqrt{\frac{1+(q_0R)^2}{1+(qR)^2}}$~\cite{blatt-weisskopf}, $q_0$
is daughter momentum calculated assuming pole mass of its mother,
$R=1.6\,\gev^{-1}$; $\Gamma(s)$ is the energy-dependent width,
$\Gamma(s)=\Gamma(\frac{q}{q_0})^3\frac{M}{\sqrt{s}}F^2$.
The function $f$ is convolved with the detector resolution function,
is integrated over the $10\,\mevm$ wide bin and is corrected for
reconstruction efficiency. The detector resolution is parameterized by
a single Gaussian function with $\sigma=5.2\,\mevm$ as determined from
MC simulation. The result of the fit is shown in
Fig.~\ref{nhb_n2s_vs_mmp} and is summarized in
Table~\ref{tab:results}. The non-resonant amplitude is found to be
consistent with zero, $b=0.03\pm0.04$. The confidence level of the fit
is 81\%. We find that the hypothesis of two resonances is favored over
the hypothesis of a single resonance (no resonances) at the
$7.4\,\sigma$ ($17.9\,\sigma$) level.

The systematic uncertainty is estimated by varying (within $\pm2$) the
order of the Chebyshev polynomial and repeating the fit to the
data. We also perform fits with the constraint on the normalization of
the $\Uth\to\Uo\pp$ reflection released.  We shift the $\mmp$ binning
by half the bin-size.  To estimate the uncertainty due to fit model,
we try to fit the $\mmp$ spectrum with the non-resonant component
fixed at zero. We vary the Blatt-Weisskopf parameter $R$ from $0$ to
$5\,\gev^{-1}$. As found in Ref.~\cite{Belle_hb}, the detector
resolution as determined from a MC simulation could be underestimated
by as much as 5-10\%. To account for this effect we repeat the fit to
the $\mmp$ spectrum with a corrected resolution function. The maximal
change of signal parameters for each source is considered as a
systematic uncertainty. We find that the systematic uncertainties
associated with releasing the constraint on the $\Uth\to\Uo\pp$
reflection and varying $R$ are negligible.  Finally, we find about
$1\,\mevm$ deviations of the $\Uo$, $\Ut$ and $\Uth$ peak positions
relative to the PDG values when the $\Un$ states are reconstructed
inclusively~\cite{Belle_hb}. These small deviations could be due to the
local variations of the background shape that are not fully described by the
polynomial. Consequently, we add an additional
uncertainty of $1\,\mevm$ to all mass measurements. All contributions
are added in quadrature to obtain the total systematic
uncertainty. The summary of systematic uncertainties study is
presented in Table~\ref{tab_syst}. The minimal level at which the
two-resonance hypothesis is favored over the one-resonance
(no-resonance) hypothesis for all considered variations is
$6.6\,\sigma$ ($16.0\,\sigma$).

The analysis of the $\hbp\,\pp$ final state follows the same strategy
as that for the $\hb\,\pp$. In this case we require
$\mmp>10.57\,\gevm$ to avoid double counting of signal events. We also
require $\mmpp<10.34\,\gevm$ and $\mmp<10.7\,\gevm$ to minimize
influence of the reflections from events where the two charged pions
originate from a $\ks\to\pp$ decay (see Fig.~\ref{mc_hb_signal}).  
We subdivide the interval $10.57\,\gevm<\mmp<10.7\,\gevm$ into 13 bins
and perform $\chi^2$ fits to the $\mmpp$ spectra for each $\mmp$
bin. The fit function consists of three components: the $\hbp$ signal,
reflection from the $\Ut\to\Uo\pp$ decay and combinatorial
background. The $\Uth$ signal is outside the mass range. The shapes of
the signal and reflection are determined from the analysis of the
exclusive $\Uf\to\uu\pp$ data.
The yield of the reflection for each $\mmp$ bin is determined from the
exclusive $\uu\pp$ data and is normalized to the ratio of inclusive and
exclusive yields. The systematic uncertainty in the inclusive yield
estimated in Ref.~\cite{Belle_hb} is taken into account. 
When fitting the inclusive $\mmpp$ spectra in bins of $\mmp$ we allow 
the yield of the reflection to float within the uncertainty of the above
measurement.
Combinatorial background is parameterized by a Chebyshev polynomial function.
The order of the polynomial varies from 6 to 8 for different $\mmp$ bins.
The fit results for the $\hbp$ yield is presented in
Fig.~\ref{nhb_n2s_vs_mmp}.

\begin{table}[!t]
\caption{Systematic uncertainties in the mass and width of the $\zbo$
  \mbox{(index 1)} and $\zbt$ (index 2), their relative normalization factor
  $a$ and phase $\phi$ for the $\hb\pp/\hbp\pp$ decay modes.}
\label{tab_syst}
\renewcommand{\arraystretch}{1.1}
\begin{ruledtabular}
\begin{tabular}{l|ccccc}
& Chebyshev        & $\mmp$  & Fit   & Resolution & Total \\
& polynomial order & binning & model &            &       \\
\hline
$M_1$, $\mevm$     & $^{+2.6}_{-0}$/$^{+4}_{-1}$ & $^{+0.1}_{-0}$/$^{+0}_{-2}$ & $^{+1.1}_{-0}$/$^{+3}_{-0}$ & $^{+0.3}_{-0}$/$^{+0}_{-0}$ & $^{+3.0}_{-1.0}$/$^{+5}_{-2}$ \\
$\Gamma_1$, $\mev$ & $^{+1.5}_{-0}$/$^{+9}_{-3}$ & $^{+1.4}_{-0}$/$^{+10}_{-0}$ & $^{+0.1}_{-0}$/$^{+0}_{-3}$ & $^{+0}_{-1.2}$/$^{+0}_{-1}$ & $^{+2.1}_{-1.2}$/$^{+13}_{-4}$ \\
$M_2$, $\mevm$     & $^{+0}_{-0.4}$/$^{+1}_{-2}$ & $^{+0}_{-0.6}$/$^{+2}_{-0}$ & $^{+0}_{-1.4}$/$^{+0}_{-0}$ & $^{+0}_{-0.3}$/$^{+0}_{-0}$ & $^{+1.0}_{-1.9}$/$^{+2}_{-2}$ \\
$\Gamma_2$, $\mev$ & $^{+0}_{-5.5}$/$^{+2}_{-2}$ & $^{+2.1}_{-0}$/$^{+4}_{-0}$ & $^{+0}_{-0.2}$/$^{+7}_{-0}$ & $^{+0}_{-0.8}$/$^{+0}_{-1}$ & $^{+2.1}_{-5.7}$/$^{+8}_{-2}$ \\
$a$                & $^{+0.1}_{-0.5}$/$^{+0.2}_{-0.7}$ & $^{+0}_{-0}$/$^{+0}_{-0.1}$ & $^{+0}_{-0.1}$/$^{+0.4}_{-0}$ & $^{+0}_{-0}$/$^{+0}_{-0}$ & $^{+0.1}_{-0.5}$/$^{+0.4}_{-0.7}$ \\
$\phi$, degree     & $^{+4}_{-4}$/$^{+11}_{-167}$ & $^{+0}_{-2}$/$^{+5}_{-0}$ & $^{+0}_{-8}$/$^{+0}_{-74}$ & $^{+0}_{-1}$/$^{+0}_{-0}$ & $^{+4}_{-9}$/$^{+12}_{-193}$ \\
\end{tabular}
\end{ruledtabular}
\end{table}

To fit the $\mmp$ distribution of the $\hbp$ signal we use the same fit
function as for the $\hb$. The result of the fit is shown in
Fig.~\ref{nhb_n2s_vs_mmp}. As in the case of the $\hb$, the
non-resonant amplitude is found to be consistent with zero,
$0.11\pm0.13$. Numerical values for other parameters are given in
Table~\ref{tab:results}. The confidence level of the fit
is 42\%. The two-resonance hypothesis is favored over that for one resonance
(no resonances) at the $2.7\,\sigma$ ($6.3\,\sigma$) level.

The systematic uncertainty is estimated in the same way as for the
$\hb$, the results are given in Table~\ref{tab_syst}. The minimal level at
which the two-resonance hypothesis is favored over the one-resonance
(no-resonance) one for all considered variations is 
$1.9\,\sigma$ ($4.7\,\sigma$).

\section{\boldmath Angular analysis}

We perform angular analyses to check consistency of the $J^P=1^+$
assignment for the $\zbo$ and $\zbt$ states and to attempt to
discriminate against other $J^P$ hypotheses; we consider $J^P=1^-$,
$2^+$ and $2^-$. The $0^+$ ($0^-$) assignment is forbidden by parity
conservation in $\zb\to\Un\pi$ ($\zb\to h_b(mP)\pi$) decays.
We use the polar angles of the pions (denoted as $\theta_1$ for a pion
from the $Y(5S)$ decay and $\theta_2$ for a pion from the $Z_b$
decay), the spatial angle, $\theta_{\pi\pi}$, between the two pions
and the angle, $\phi_p$, between the plane defined by the pion from
$Y(5S)$ decay and the beam direction and the plane defined by the two
pions.  Since the $\zb$ velocity is very small, $\beta<0.02$, we
neglect its recoil motion and measure all pion momenta in the
c.m.\ frame.  We assume that only the lowest partial wave contributes to the
decay in the cases where more than one partial wave is
possible~\cite{milstein}.

\subsection{$\Ut\pp$ and $\Uth\pp$ final states}

We apply an additional requirement on the angle between the two pion
momenta measured in the laboratory frame, $\cos\theta_{\rm lab}<0.95$,
to suppress the background from converted photons.

\begin{figure}[!t]
\vspace*{-5mm}
\includegraphics[width=0.32\textwidth]{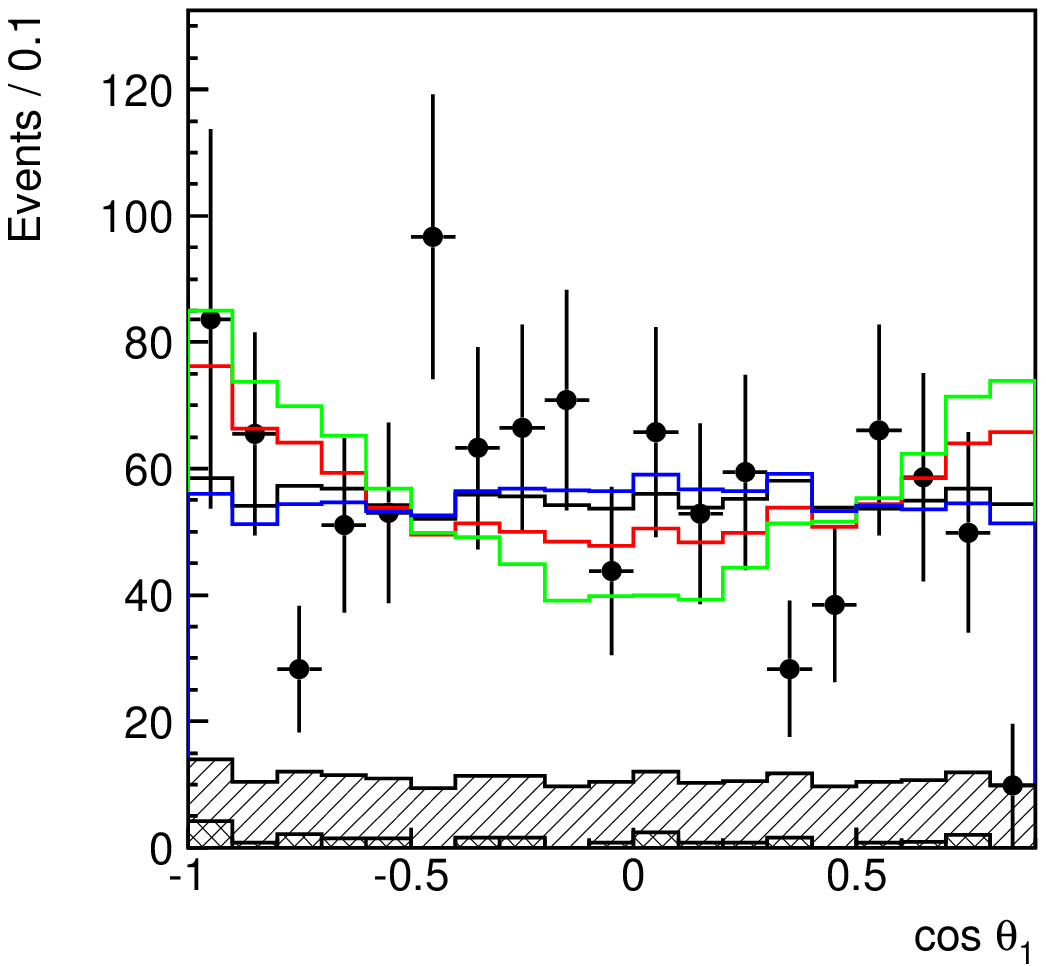}
\hfill
\includegraphics[width=0.32\textwidth]{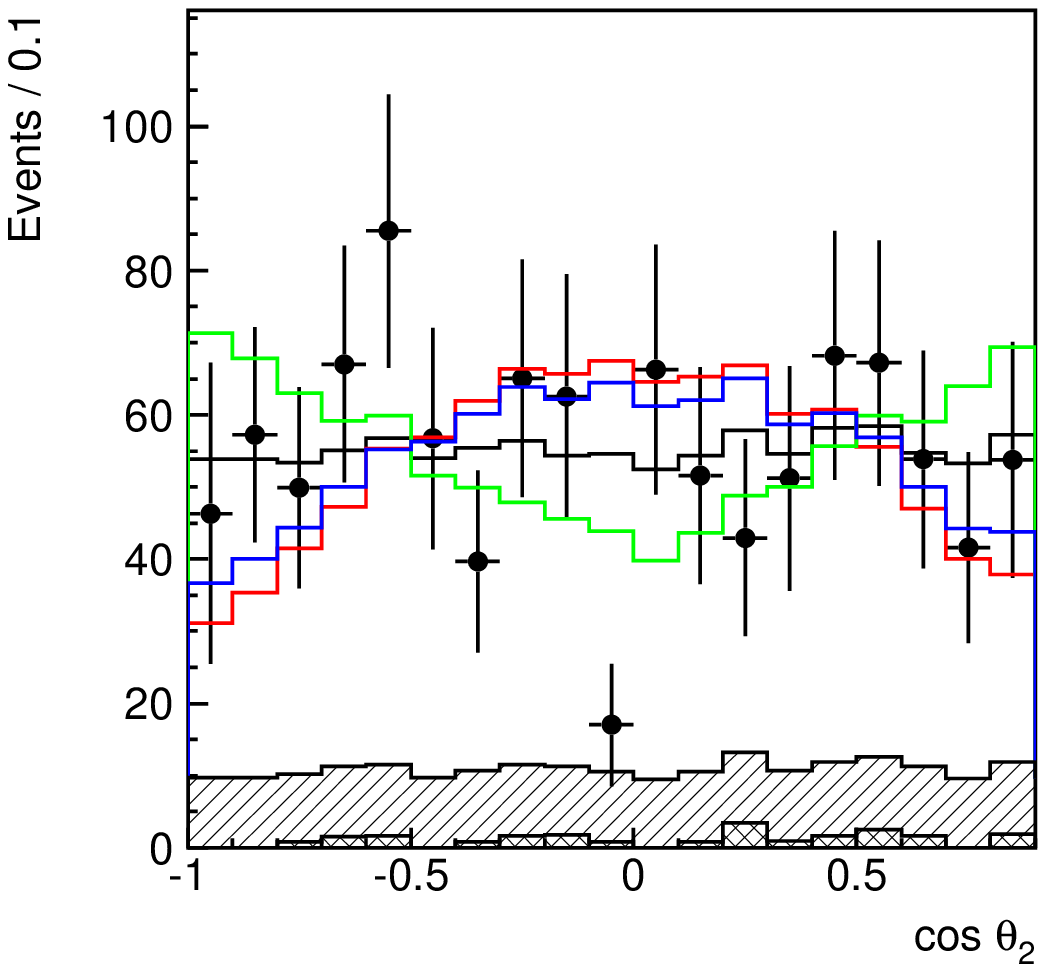}
\hfill
\includegraphics[width=0.32\textwidth]{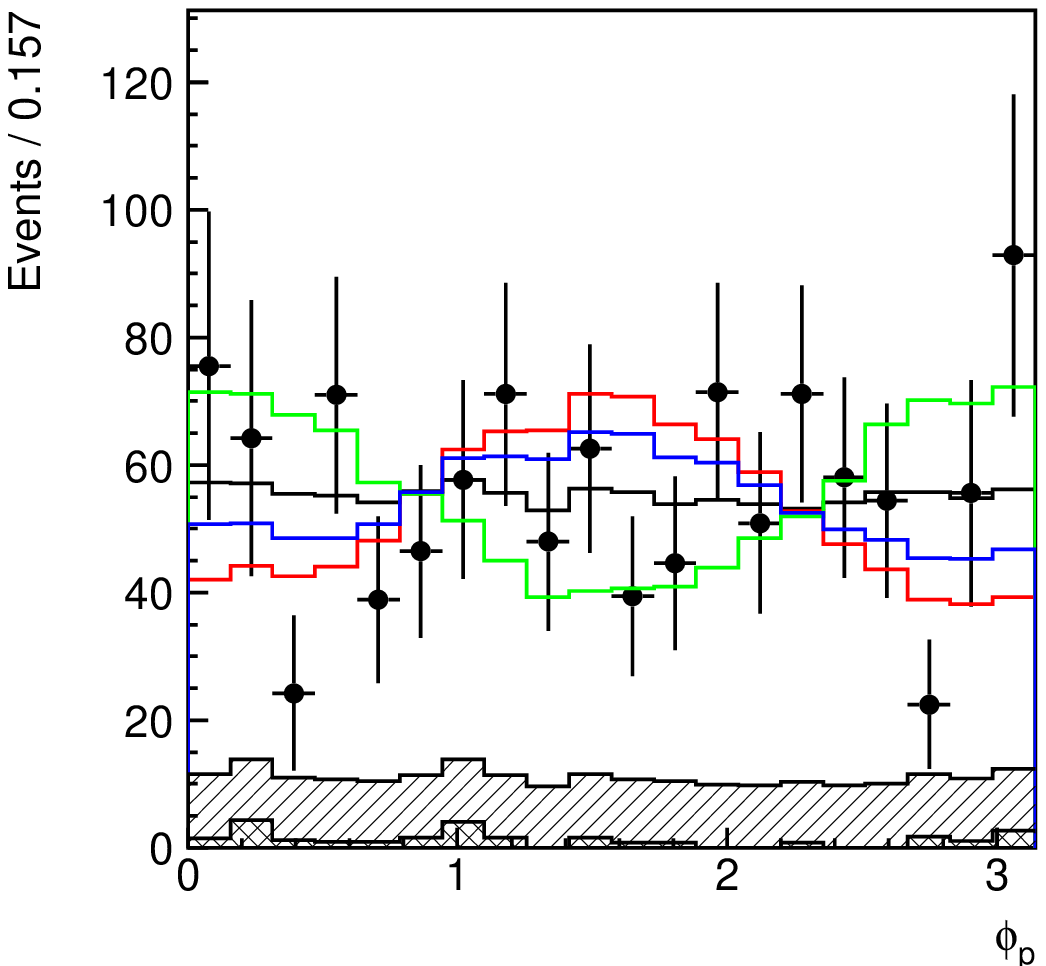} \\
\vspace*{-5mm}
\includegraphics[width=0.32\textwidth]{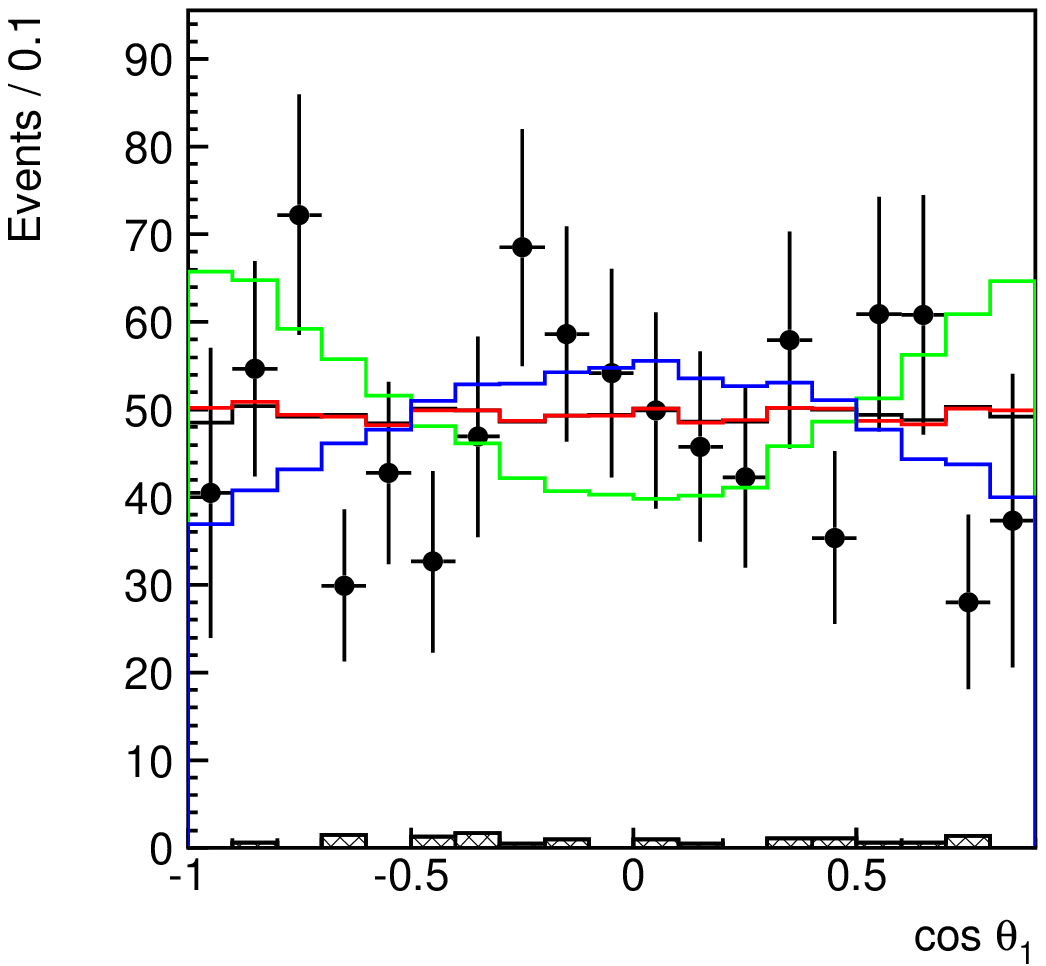}
\hfill
\includegraphics[width=0.32\textwidth]{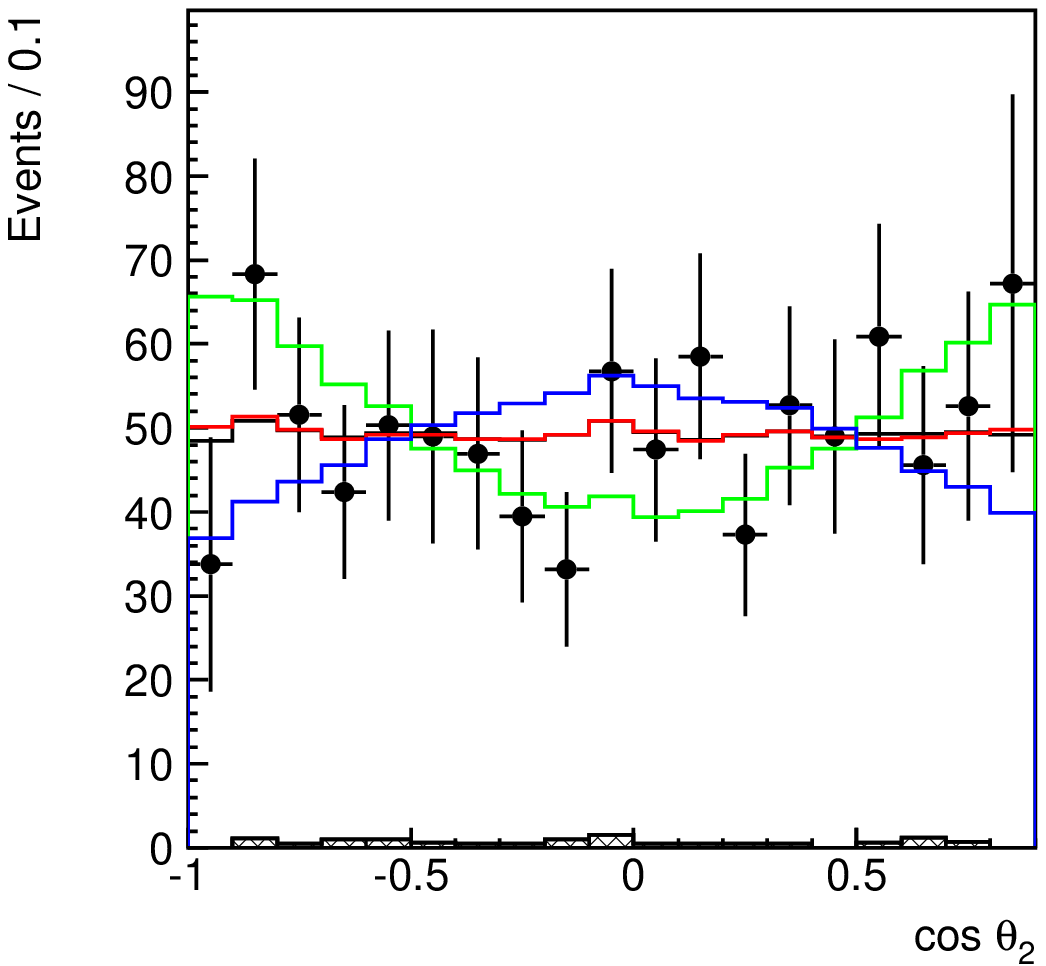}
\hfill
\includegraphics[width=0.32\textwidth]{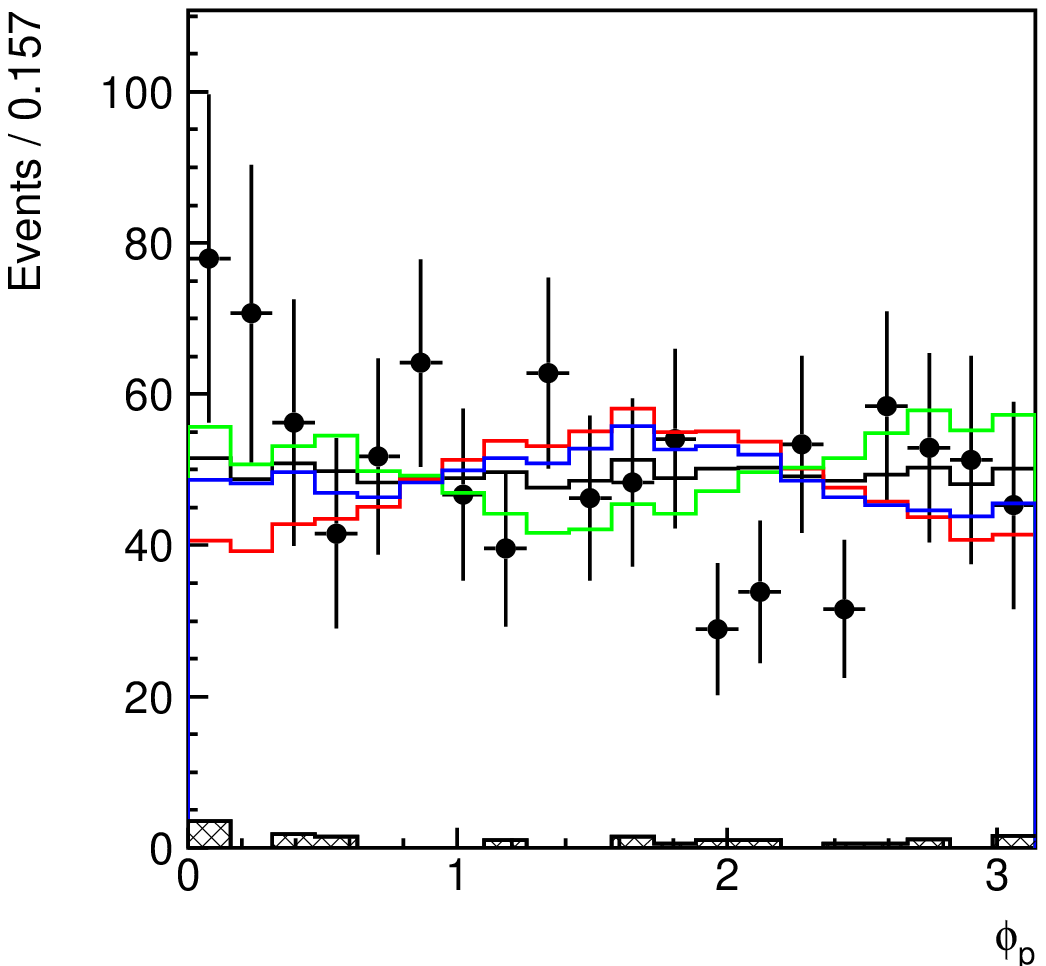}
\caption{ Angular distributions for the $\zbo$ signal region.  Points
  with error bars represent the yield of the $\Ut\pi$ candidates
  (first row) and $\Uth\pi$ candidates (second row) as a function of
  the $\cto$ (left column), $\ctt$ (middle column) and $\phi_p$ (right
  column).  The open histograms represent the fit results for
  different $J^P$ hypotheses: $1^+$ (black), $1^-$ (red), $2^+$
  (green) and $2^-$ (blue).  The hatched (cross-hatched) histogram
  shows the contribution of the non-resonant (combinatorial)
  component. }
\label{cto}
\end{figure}

\begin{figure}[!t]
\vspace*{-5mm}
\includegraphics[width=0.32\textwidth]{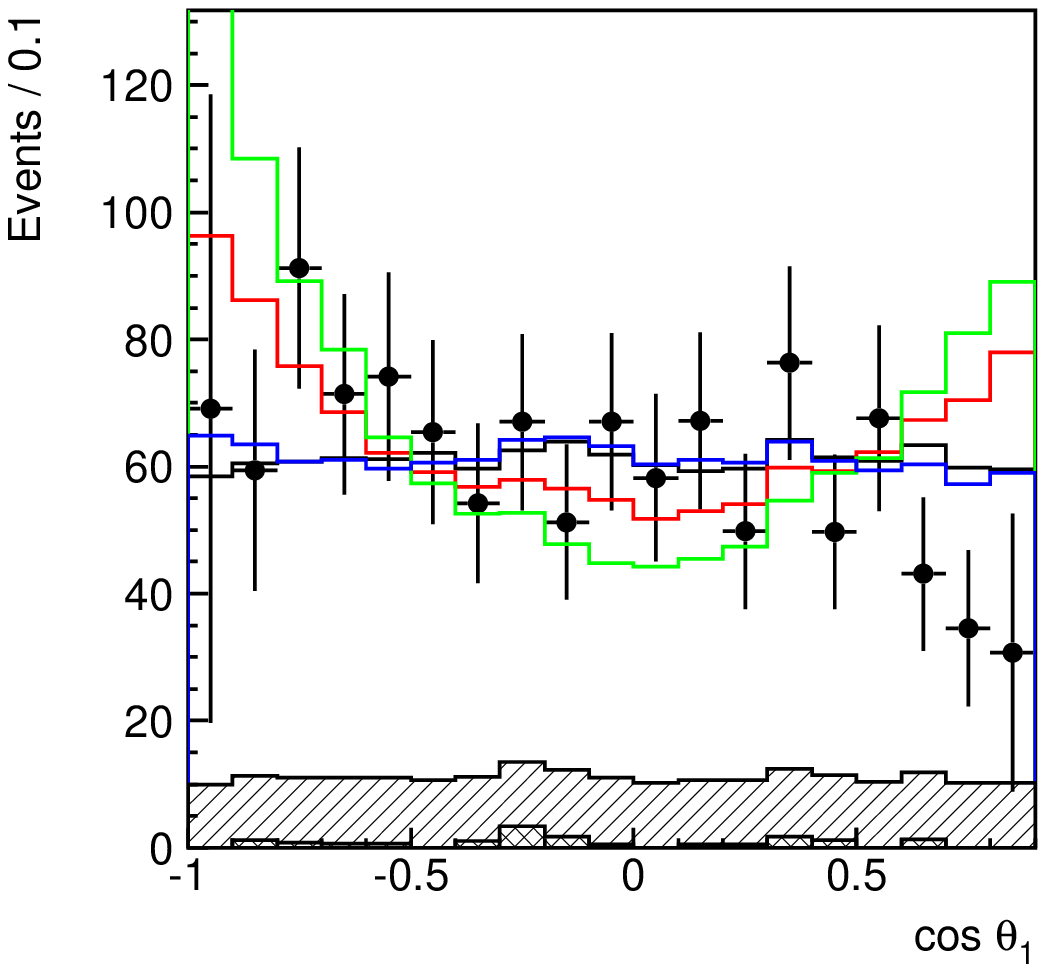}
\includegraphics[width=0.32\textwidth]{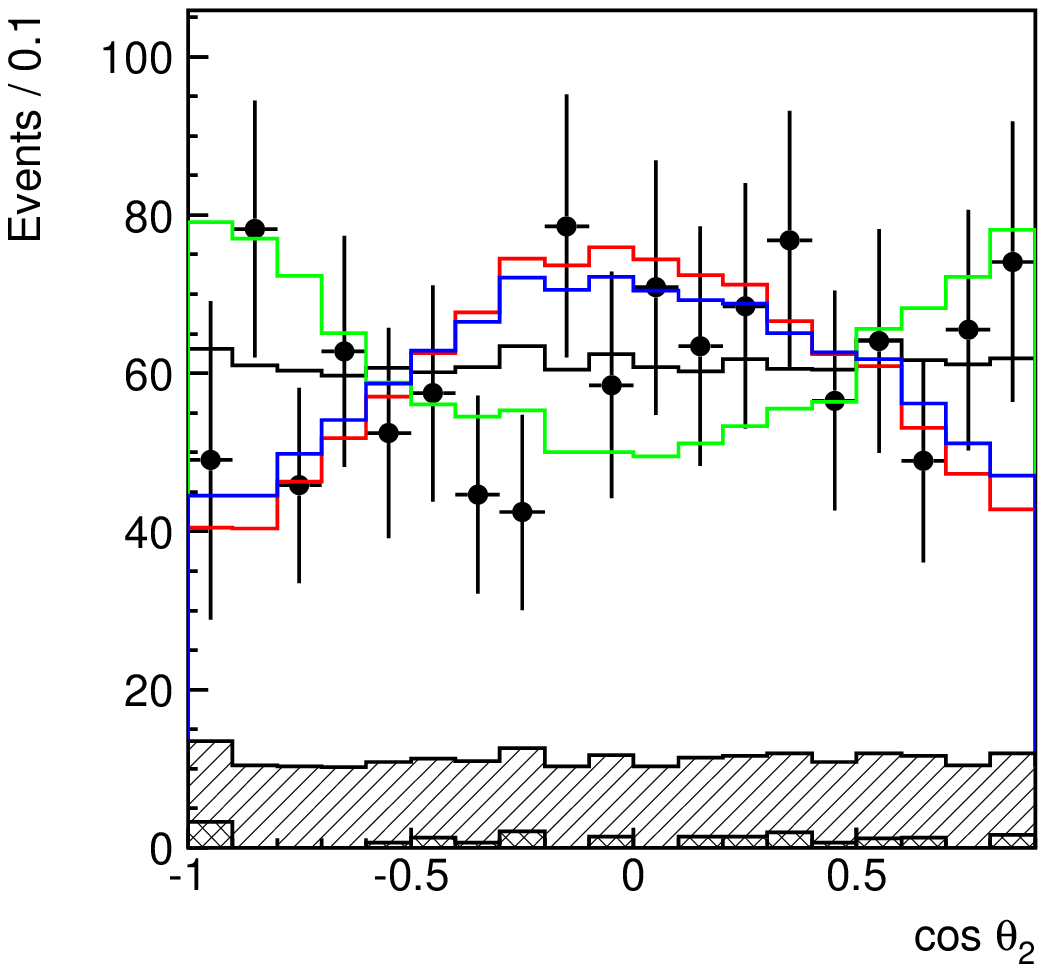}
\includegraphics[width=0.32\textwidth]{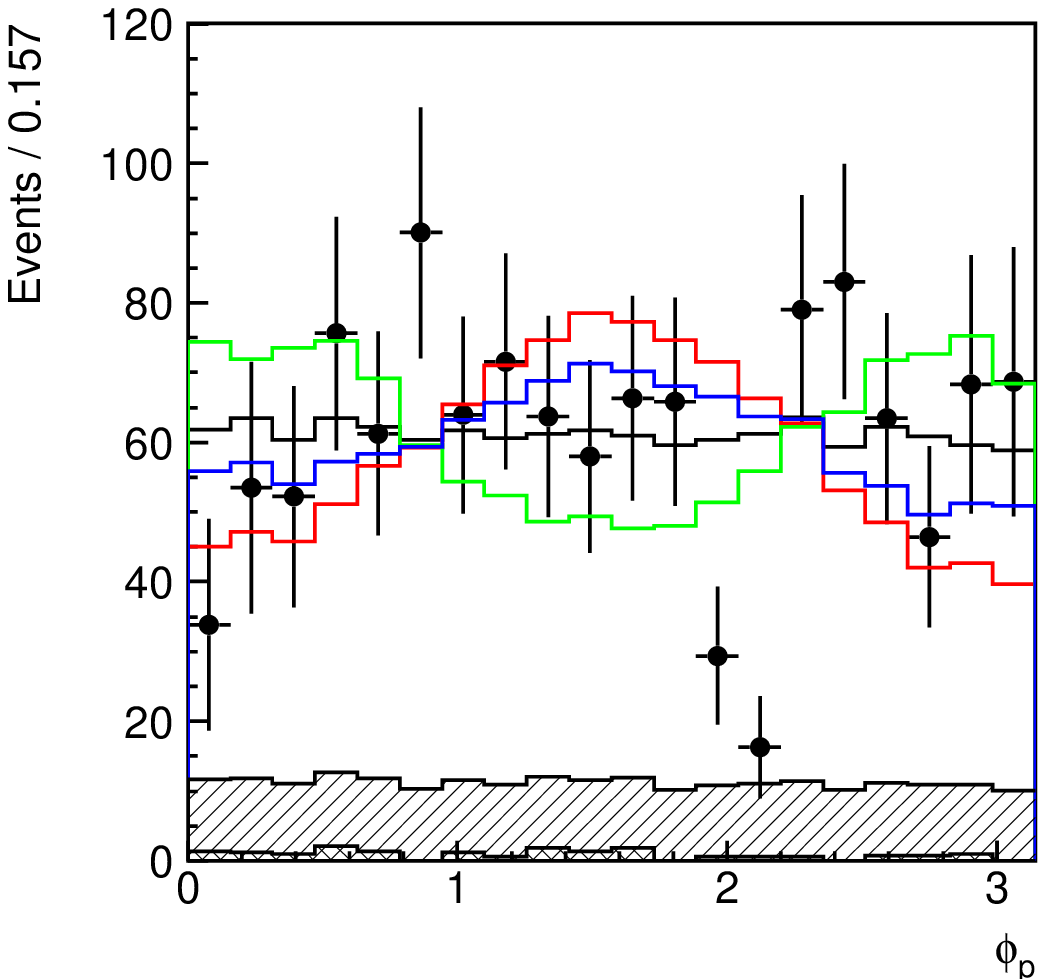} \\
\vspace*{-5mm}
\includegraphics[width=0.32\textwidth]{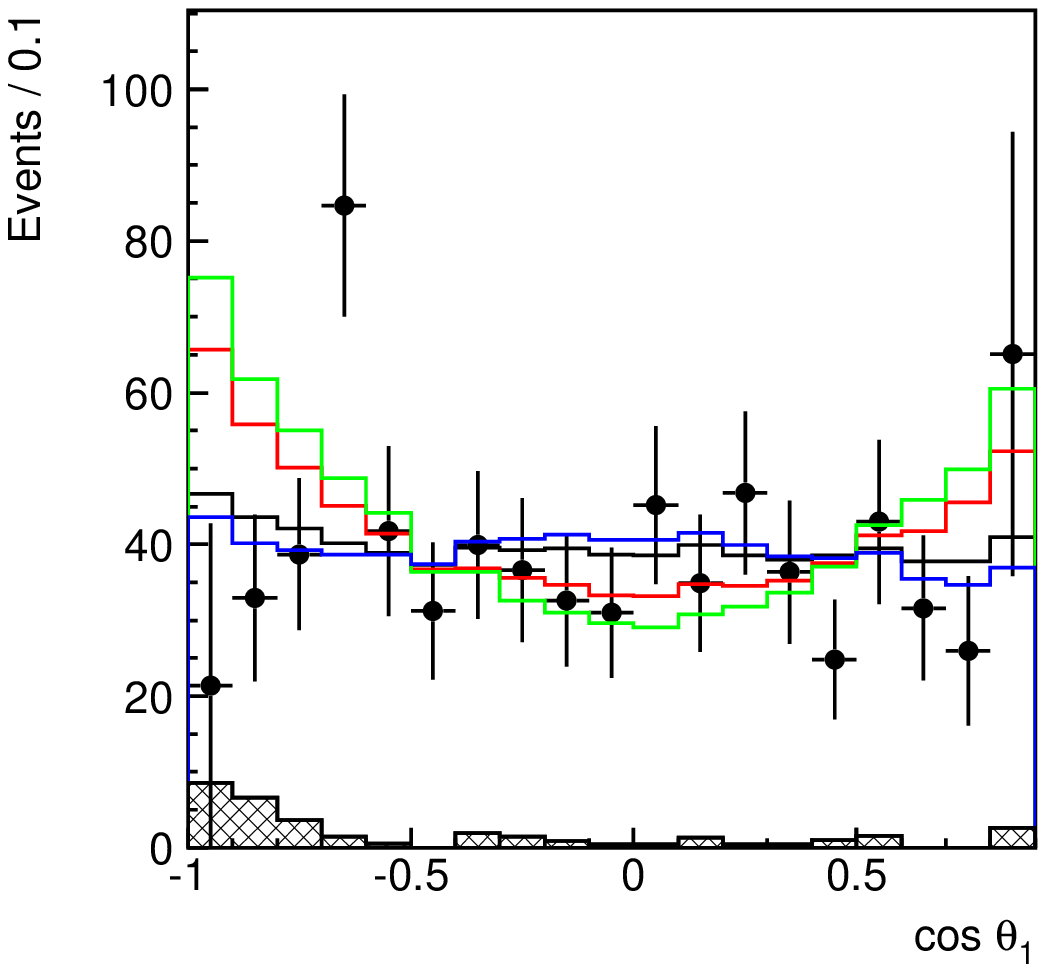}
\includegraphics[width=0.32\textwidth]{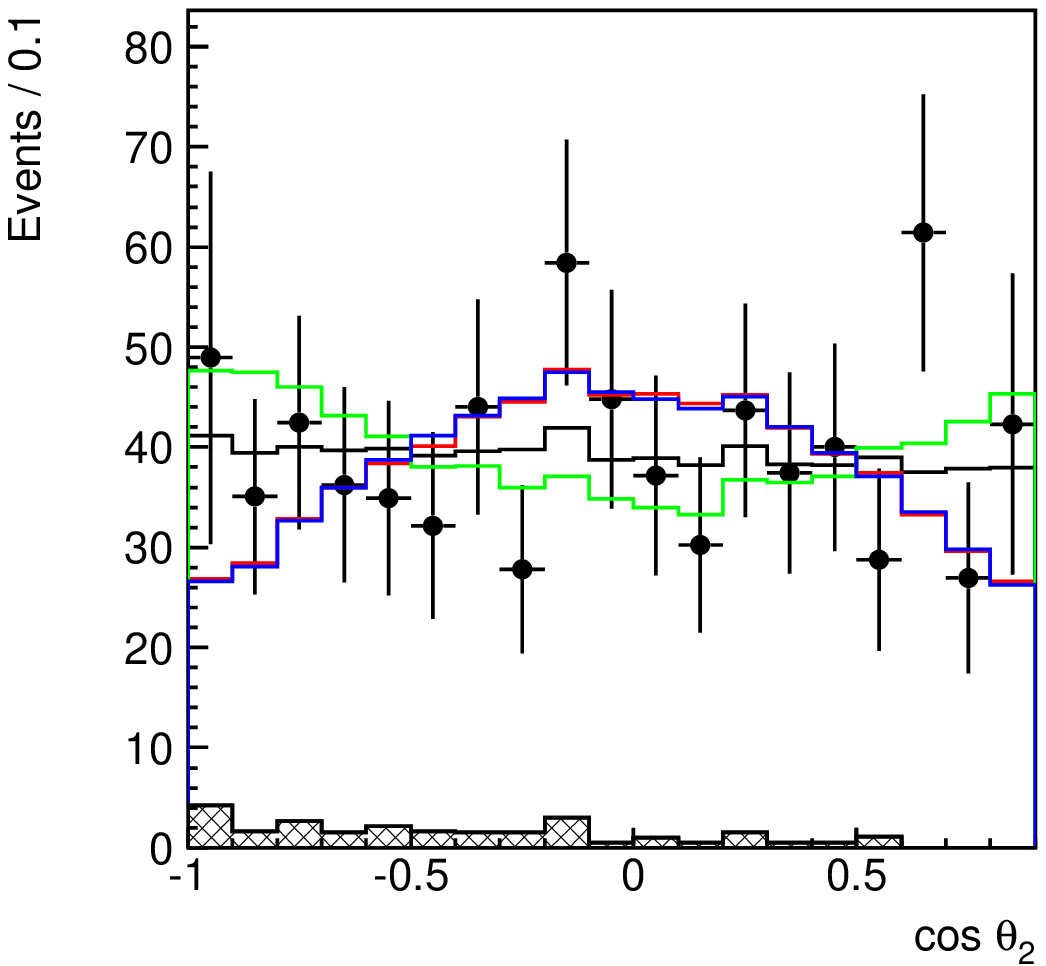}
\includegraphics[width=0.32\textwidth]{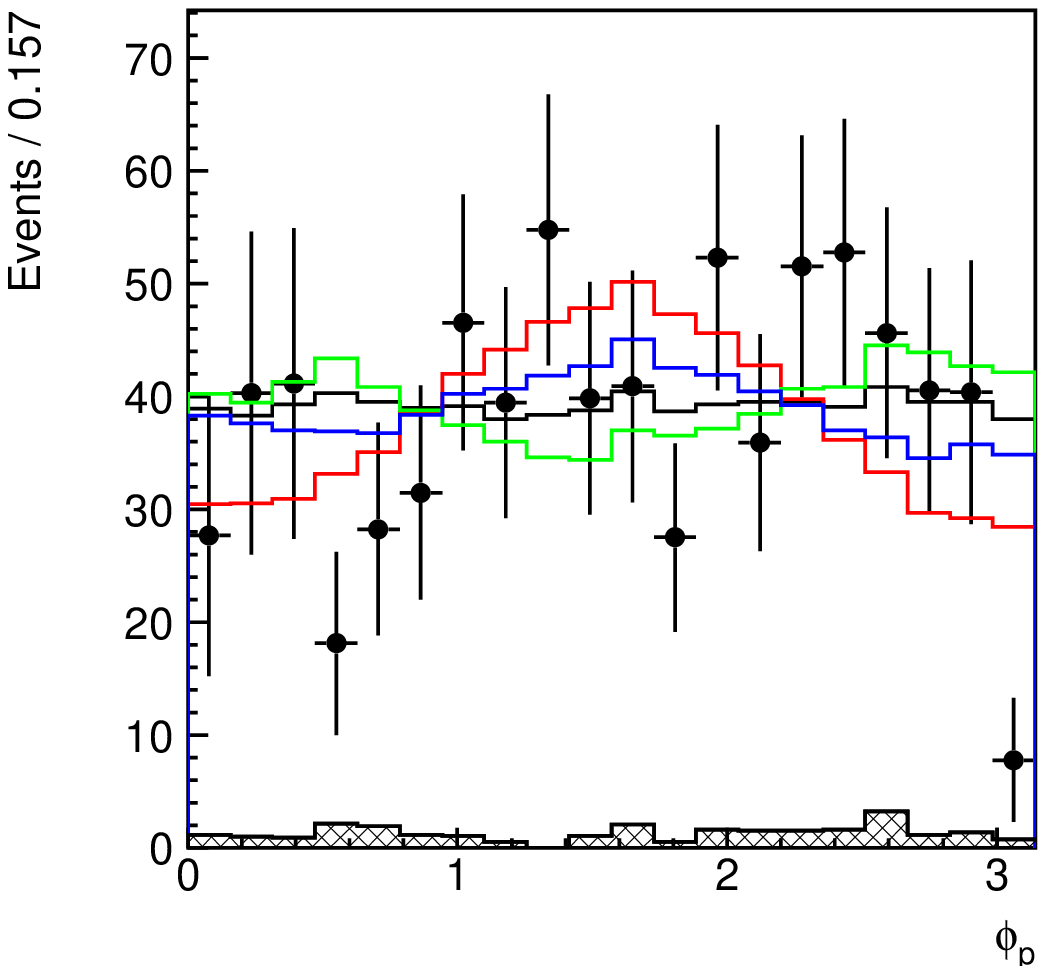}
\caption{ Angular distributions for the $\zbt$ signal region. The
  histogram contents are similar to those shown in Fig.~\ref{cto}.}
\label{cto2}
\end{figure}

We first study the angular distribution for the non-resonant component
for the $\Ut\pp$ candidates. The non-resonant region is defined as
$10.5\,\gevm<M(\U\pi)<10.58\,\gevm$. The $\ctpp$ values is tightly
correlated with the $\mpp$, therefore the $\ctpp$ distribution
reflects the details of the $\mpp$ distribution discussed in the
Dalitz analysis section of the paper.  The $\cto$, $\ctt$ and $\phi_p$
distributions (after the $\mmpp$ sideband subtraction and efficiency
correction) are all consistent with uniform distributions. This
indicates that the non-resonant contribution is dominated by an
$S$-wave component.

At the next step we select the $\zbo$ [$\zbt$] region
$10.602<M(\Ut\pi)<10.638\,\gevm$ [$10.648<M(\Ut\pi)<10.7\,\gevm$] for
the $\Ut\pp$ final state and $10.607<M(\Uth\pi)<10.627\,\gevm$
[$10.649<M(\Ut\pi)<10.676\,\gevm$] for the $\Uth\pp$ final state.  For
the $\Ut\pp$ candidates we require
$0.15\,\gevms<M^2(\pp)<0.4\,\gevms$, which considerably suppresses the
non-resonant contribution. The $\cto$, $\ctt$ and $\phi_p$
distributions corrected for the reconstruction efficiency are shown in
Figs~\ref{cto} and \ref{cto2}.

We perform a binned maximum likelihood fit to these distributions. The
fit function is a sum of three components: (a) the $\zb$ signal, (b)
the contribution due to the non-resonant component and (c)
combinatorial background described by a properly normalized $\mmpp$
sidebands.  The terms due to the interference of the $\zb$ signal and
$S$-wave non-resonant part are constant in $\cto$, $\ctt$ and
$\phi_p$. Therefore, component (b) of the fit function is a
constant. We estimate its normalization from the linear extrapolation
in the $M(\Ut\pi)$ for the $\Ut\pp$ final state and we assume that it
is zero for the $\Uth\pp$ final state.  The non-resonant contribution
varies with $\ctpp$, therefore this variable is not used.  The
components (a) and (b) of the fit function are corrected for
efficiency. The only floating parameter in the fit is the
normalization of the $\zb$ signal component~\cite{angular_Y3S_Zb1}. 

The fit results for various $\zb$ spin-parity assignments are shown in
Figs.~\ref{cto} and \ref{cto2}.  We use $\ctt$ to discriminate $1^+$
from $1^-$ and $1^+$ from $2^-$, and we use $\cto$ to discriminate
$1^+$ from $2^+$.  The probabilities at which the $1^-$, $2^+$ and
$2^-$ hypotheses are disfavored compared to the $1^+$ hypothesis are
calculated as $\sqrt{\Delta2\log{L}}$ and given in
Table~\ref{2logL_cto}.

\begin{table}[!b]
\caption{ The probabilities at which different $J^P$ hypotheses are
  disfavored compared to the $1^+$ hypothesis. The data are consistent
  with the $1^+$ hypothesis. }
\label{2logL_cto}
\begin{tabular}{|c|c|c|c|c|c|c|} \hline
\multirow{2}{*}{\;$J^P$\;}& \multicolumn{3}{|c|}{$\zbo$} & \multicolumn{3}{|c|}{$\zbt$} \\ \cline{2-7}
& \;$\Ut\pp$\; & \;$\Uth\pp$\; & \;$\hb\pp$\; & \;$\Ut\pp$\; & \;$\Uth\pp$\; & \;$\hb\pp$\; \\ \hline
$1^-$ & $3.6\,\sigma$ & $0.3\,\sigma$ & $0.3\,\sigma$ & $3.7\,\sigma$ & $2.6\,\sigma$ & $2.7\,\sigma$ \\ \hline
$2^+$ & $4.3\,\sigma$ & $3.5\,\sigma$ & \multirow{2}{*}{$4.3\,\sigma$} & $4.4\,\sigma$ & $2.7\,\sigma$ & \multirow{2}{*}{$2.1\,\sigma$} \\ \cline{1-3} \cline{5-6}
$2^-$ & $2.7\,\sigma$ & $2.8\,\sigma$ &  & $2.9\,\sigma$ & $2.6\,\sigma$ &  \\ \hline
\end{tabular}
\end{table}

\begin{figure}[!t]
\includegraphics[width=0.32\textwidth]{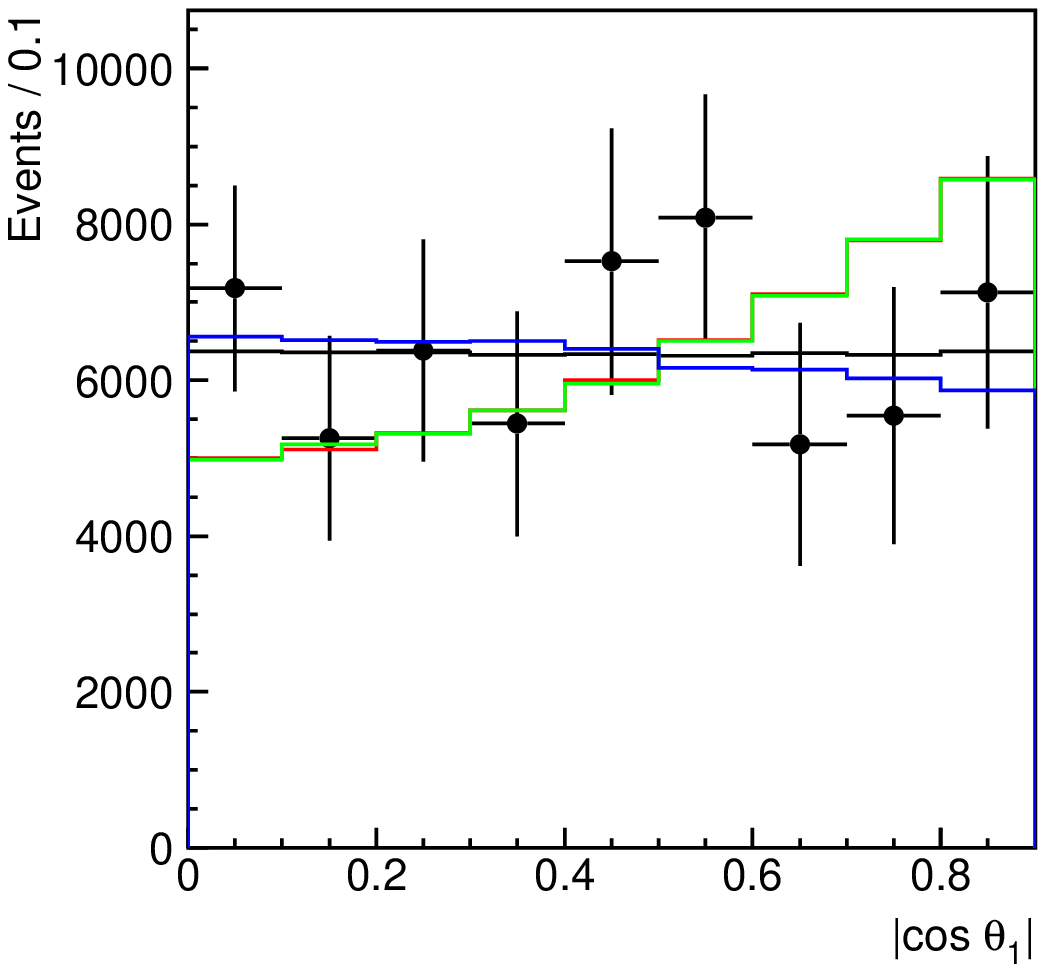}
\includegraphics[width=0.32\textwidth]{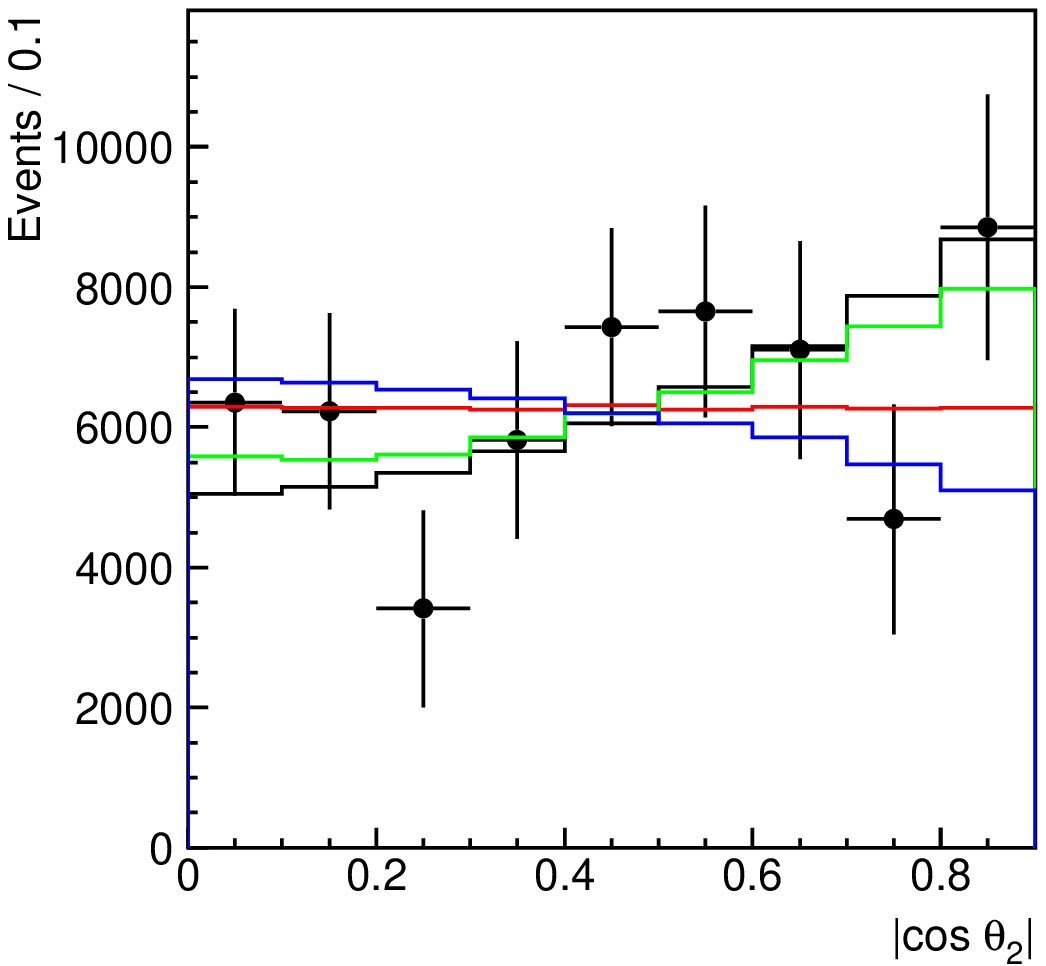}
\includegraphics[width=0.32\textwidth]{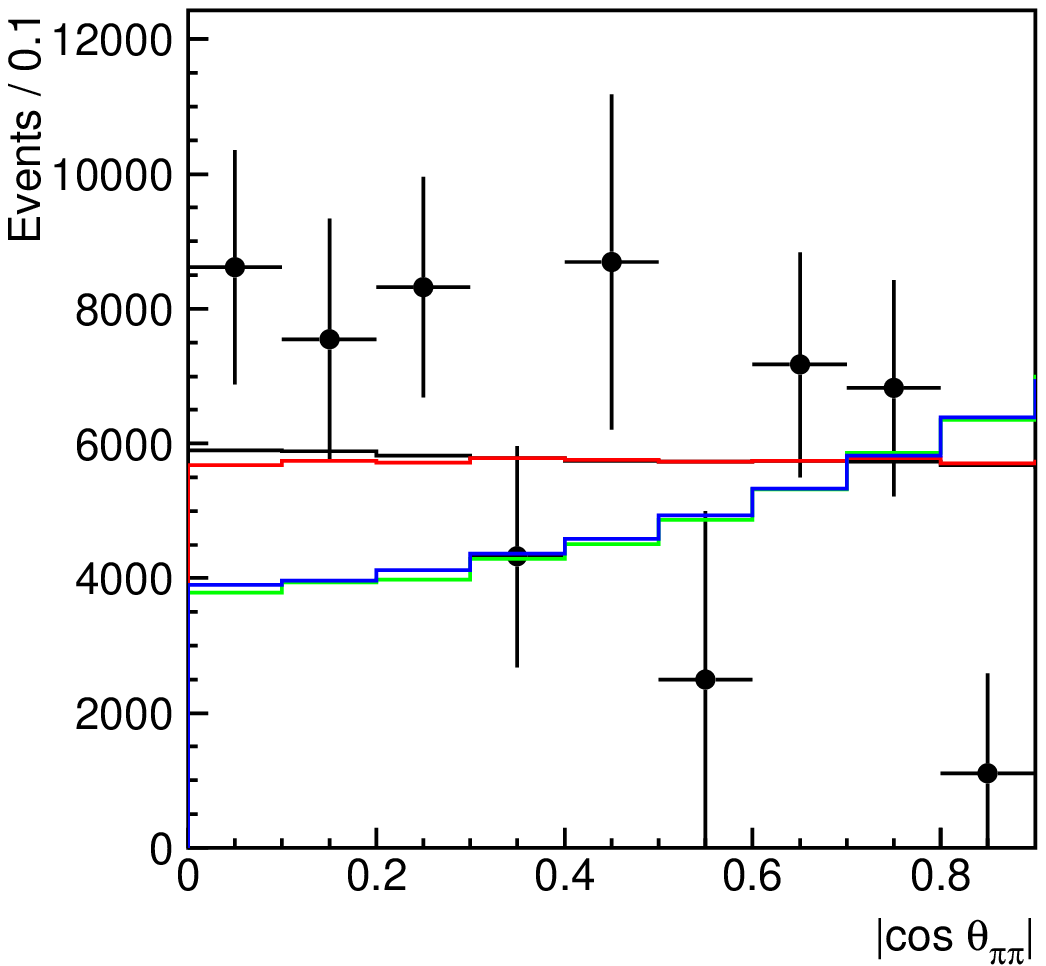}
\includegraphics[width=0.32\textwidth]{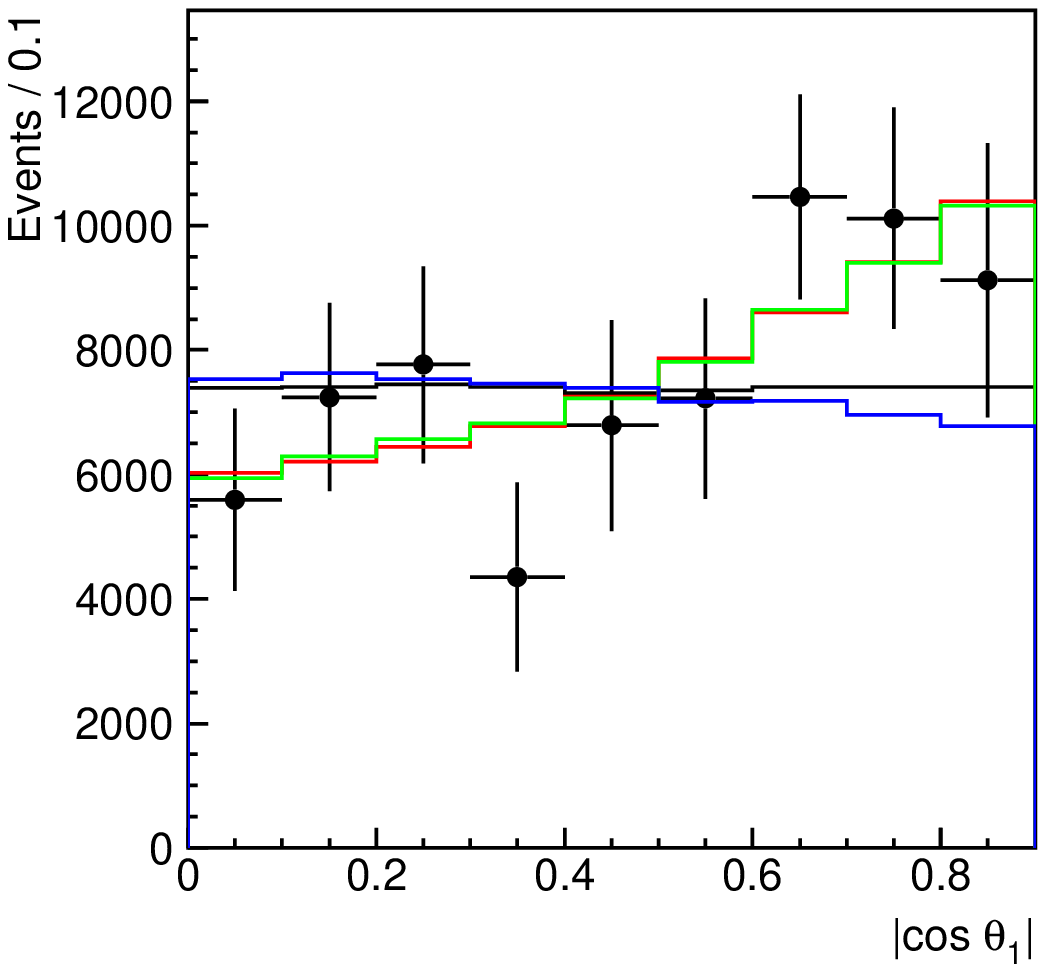}
\includegraphics[width=0.32\textwidth]{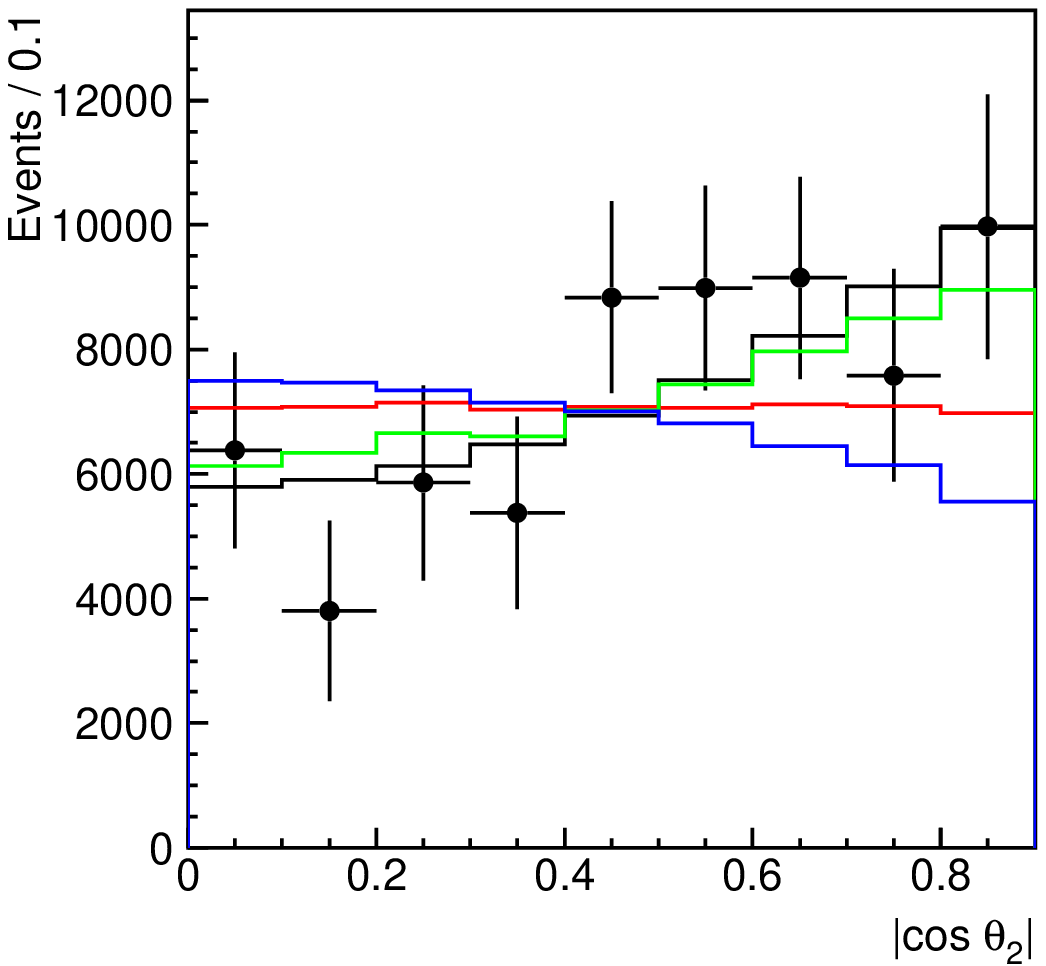}
\includegraphics[width=0.32\textwidth]{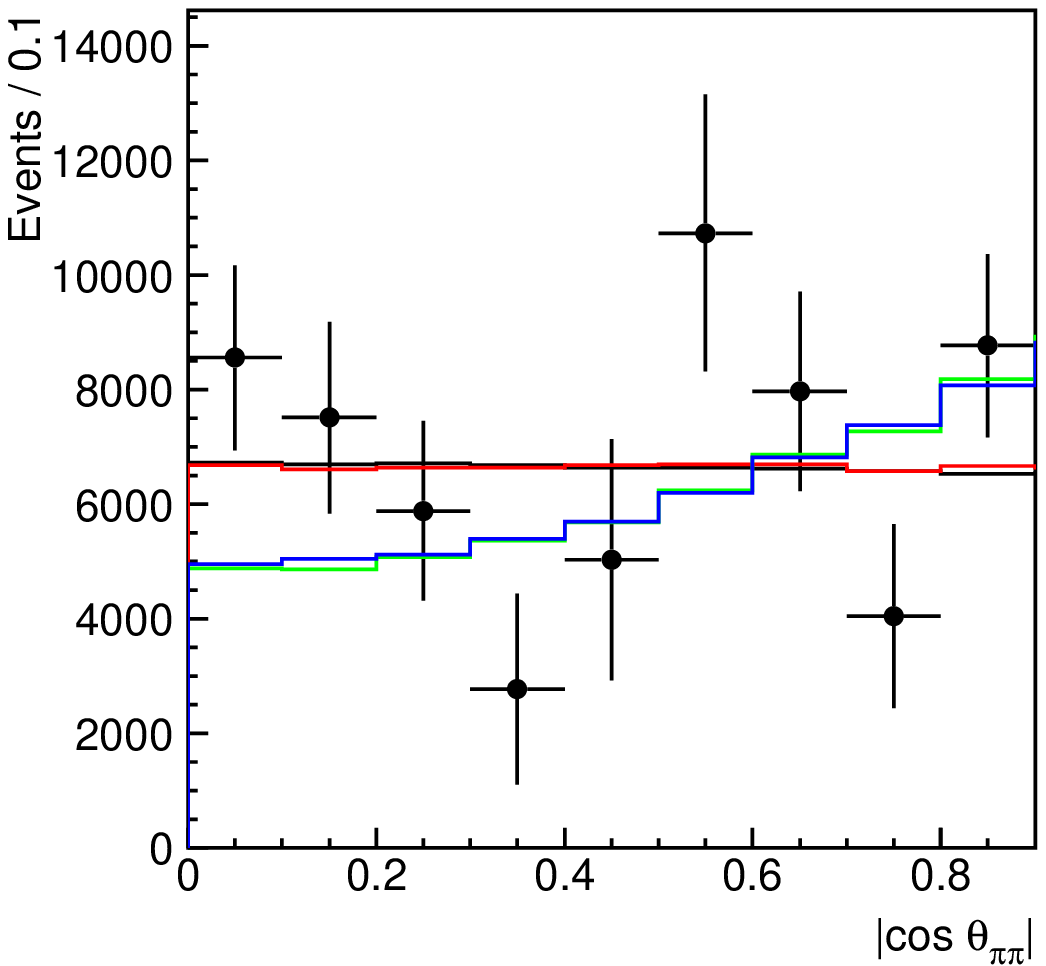}
\caption{ Points with error bars represent the $\hb$ yield as a
  function of the $\cto$ (left column), $\ctt$ (middle column) and
  $\ctpp$ (right column) for the $\zbo$ (top row) and $\zbt$ (bottom
  row) regions. Histograms represent the fit results for different
  $J^P$ hypotheses: $1^+$ (black), $1^-$ (red), $2^+$ (green) and
  $2^-$ (blue). }
\label{ct_hb}
\end{figure}

\subsection{$\hb\pp$ final state}

We define the $\zbo$ and $\zbt$ signal regions as
$10.6\,\gevm<MM(\pi)<10.63\,\gevm$ and
$10.63\,\gevm<\mmp<10.67\,\gevm$, respectively. For each region we fit
the $\mmpp$ spectra in bins of $\cto$, $\ctt$ and $\ctpp$ to determine
the $\hb$ signal yield.  The $\hb$ yield corrected for the
reconstruction efficiency as a function of angular variables is shown
in Fig~\ref{ct_hb}. The results of the fits to the different
spin-parity hypotheses are superimposed. We use $\ctpp$ to
discriminate $1^+$ from $2^+$ and $1^+$ from $2^-$, and we use $\ctt$
to discriminate $1^+$ from $1^-$.  The level at which the $1^-$, $2^+$
and $2^-$ hypotheses are disfavored compared to the $1^+$ hypothesis
is calculated as a square root of the difference of the corresponding
$\chi^2$ values and is shown in Table~\ref{2logL_cto}.

The values quoted in Table~\ref{2logL_cto} are
preliminary. Our procedure to deal with the non-resonant contribution is
approximate, and we do not take into account mutual cross-feed of the
$\zbo$ and $\zbt$ states. However, the angular analyses indicate that
the $1^+$ hypothesis for both $\zbo$ and $\zbt$ provides the best 
description of angular distributions compared to all other hypotheses with 
$J\leq2$.

\begin{table}[!t]
  \caption{Comparison of results on $Z_b(10610)$ and $Z_b(10650)$ parameters
           obtained from $\Uf\to\Un\pp$ ($n=1,2,3$) and $\Uf\to \hbn\pp$ 
           ($m=1,2$) analyses. Quoted values are in MeV/$c^2$ for masses, in
           MeV for widths and in degrees for the relative phase. Relative 
           amplitude is defined as $a_{Z_b(10650)}/a_{Z_b{10610}}$.}
  \medskip
  \label{tab:results}
\centering
  \begin{tabular}{lccccc} \hline \hline
 Final state & $\Uo\pp$                   &
               $\Ut\pp$                   &
               $\Uth\pp$                  &
               $\hb\pp$                   &
               $\hbp\pp$
\\ \hline
           $M(Z_b(10610))$ &
           $10609\pm3\pm2$                &
           $10616\pm2^{+3}_{-4}$          &
           $10608\pm2^{+5}_{-2}$          &
           $\mzahb$          &
           $\mzahbp$ 
 \\
           $\Gamma(Z_b(10610))$ &
           $22.9\pm7.3\pm2$               &
           $21.1\pm4^{+2}_{-3}$           &
           $12.2\pm1.7\pm4$               &
           $\gzahb$   &
           $\gzahbp$ 
 \\
           $M(Z_b(10650))$ &
           $10660\pm6\pm2$                &
           $10653\pm2\pm2$                &
           $10652\pm2\pm2$                &
           $\mzbhb$      &
           $\mzbhbp$
 \\
           $\Gamma(Z_b(10650))$ &
           $12\pm10\pm3$~                 &
           $16.4\pm3.6^{+4}_{-6}$         &
           $10.9\pm2.6^{+4}_{-2}$         &  
           $\gzbhb$           & 
           $\gzbhbp$ 
 \\
           Rel. amplitude                 &
           $0.59\pm0.19^{+0.09}_{-0.03}$  &
           $0.91\pm0.11^{+0.04}_{-0.03}$  &
           $0.73\pm0.10^{+0.15}_{-0.05}$  &
           $\ahb$    &
           $\ahbp$
 \\
           Rel. phase, &
           $53\pm61^{+5}_{-50}$           &
           $-20\pm18^{+14}_{-9}$          &
           $6\pm24^{+23}_{-59}$           &
           $\phihb$         &
           $\phihbp$   
\\
\hline \hline
\end{tabular}
\end{table}

\section{Discussion and Conclusions}

\begin{figure}[!t]
\includegraphics[width=0.80\textwidth]{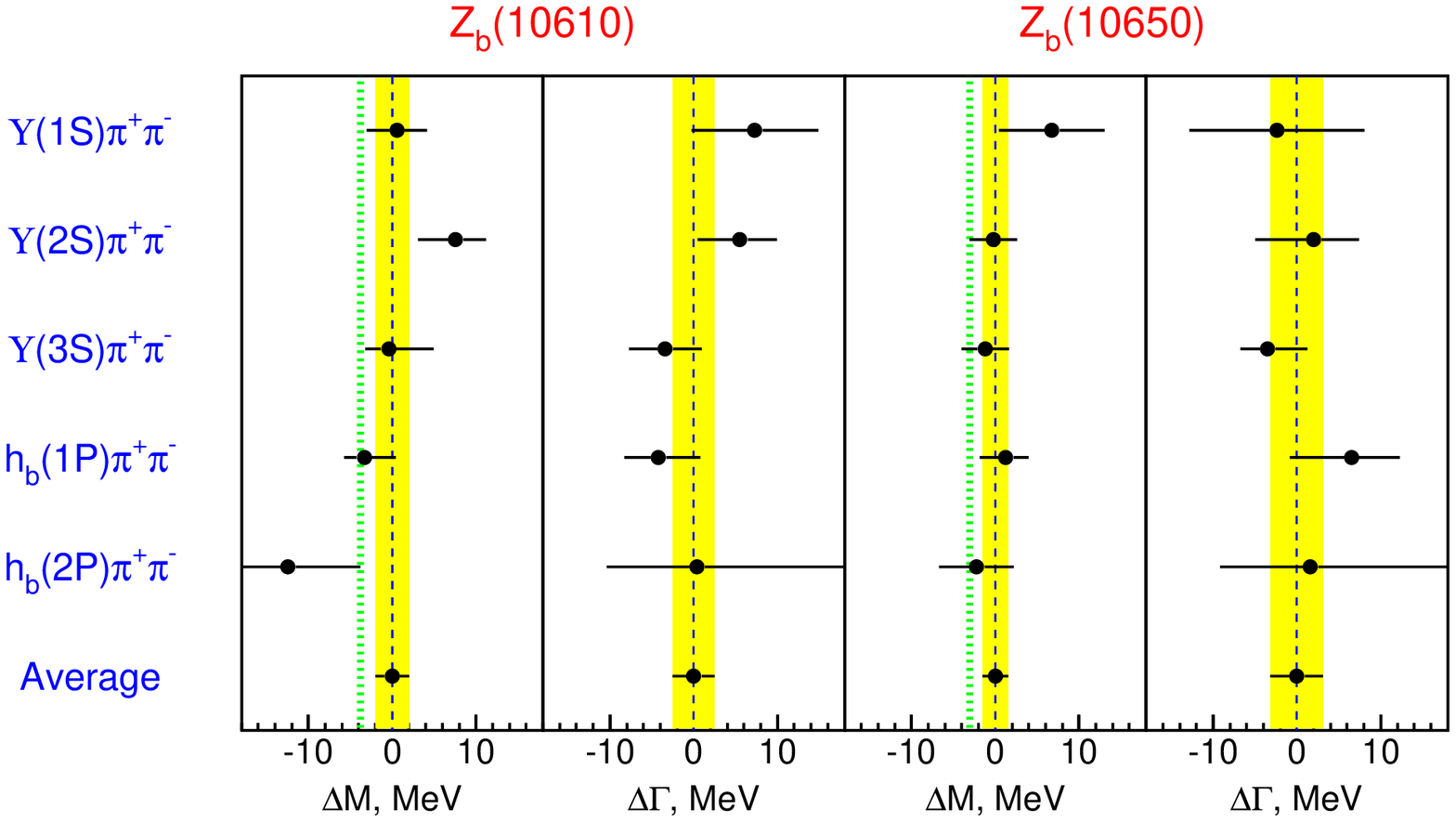}
\caption{Comparison of $Z_b(10610)$ and $Z_b(10650)$ parameters
  obtained from different decay channels. The vertical dotted lines
  indicate $B^*{\overline B}$ and $B^*{\overline B^*}$ thresholds.}
\label{fig:summary}
\end{figure}

In conclusion, we have observed two charged bottomonium-like
resonances, the $\zbo$ and $\zbt$, with signals in five different
decay channels, $\Un\pipm$ ($n=1,2,3$) and $\hbn\pipm$ ($m=1,2$).
Parameters of the resonances as measured in different channels are
summarized in Table~\ref{tab:results}. All channels yield consistent
results as can be seen in Fig.~\ref{fig:summary}.  A simple weighted
averages over all five channels give
$M[Z_b(10610)]=10608.4\pm2.0\,\mevm$,
$\Gamma[Z_b(10610)]=15.6\pm2.5\,\mev$ and
$M[Z_b(10650)]=10653.2\pm1.5\,\mevm$,
$\Gamma[Z_b(10650)]=14.4\pm3.2\,\mev$, where statistical and
systematic errors are added in quadrature.

Charged bottomonium-like resonances cannot be simple $b\bar{b}$
combinations.
The measured masses of these new states exceed by only a few MeV/$c^2$
the thresholds for the open beauty channels $B^*{\overline B}$
($10604.6$~MeV) and $B^* {\overline B^*}$ ($10650.2$~MeV).  This
``coincidence'' is suggestive of ``molecular'' states whose
their structure is determined by the strong interaction dynamics of
$B^* {\overline B}$ and $B^*{\overline B^*}$ meson pairs.

The widths of both states are similar and are of the order of
$15\,\mevm$.  The $\zbo$ production rate is similar to the $\zbt$
production rate for every decay channel. Their relative phase is
consistent with zero for the final states with the $\Un$ and
consistent with 180 degrees for the final states with $\hbn$.

The $\Uf\to\hbn\pp$ decays seem to be saturated by the $\zbo$ and
$\zbt$ intermediate states; this decay mechanism is responsible for
the high rate of the $\Uf\to\hbn\pp$ process measured recently by the
Belle Collaboration.

Analysis of angular distributions for charged pions favors the
$J^P=1^+$ spin-parity assignment for both $\zbo$ and $\zbt$.
Since the $\Upsilon(5S)$ has negative G-parity, $\zb$ states will have
opposite G-parity due to emission of the pion.

\section{Acknowledgement}

We are grateful to Alexander Milstein of BINP and Mikhail Voloshin of
the University of Minnesota for fruitful discussions.

We thank the KEKB group for excellent operation of the accelerator,
the KEK cryogenics group for efficient solenoid operations, and the
KEK computer group and the NII for valuable computing and SINET4
network support.  We acknowledge support from MEXT, JSPS and Nagoya's
TLPRC (Japan); ARC and DIISR (Australia); NSFC (China); MSMT
(Czechia); DST (India); MEST, NRF, NSDC of KISTI, and WCU (Korea);
MNiSW (Poland); MES and RFAAE (Russia); ARRS (Slovenia); SNSF
(Switzerland); NSC and MOE (Taiwan); and DOE and NSF (USA).


\begin{thebibliography}{99}
\bibitem{milstein} The angular distributions for various $J^P$
  hypothesis of the $\zb$ were calculated by A.~Milstein (private
  communications).
\bibitem{angular_Y3S_Zb1} For the $\zbo\to\Uth\pi$ decay the pion from
  the $Y(5S)$ decay and the pion from the $\zbo$ decay have similar
  momenta and are difficult to distinguish. Therefore for $\cto$
  ($\ctt$) we use cosine of the $\pip$ ($\pim$) polar angle and we
  average the fit functions among $\cto$ and $\ctt$.
\end{thebibliography}
\end{document}